\documentclass[runningheads]{llncs}

\usepackage[T1]{fontenc}
\usepackage{lmodern}
\usepackage{amsmath,amssymb}

\usepackage{xcolor}
\newif\ifshowchanges
\showchangestrue

\usepackage{microtype}
\usepackage{booktabs,tabularx}
\usepackage{tikz}
\usetikzlibrary{arrows.meta}
\usepackage[hidelinks]{hyperref}
\DeclareMathOperator{\rank}{rank}
\DeclareMathOperator{\cl}{cl}
\DeclareMathOperator{\ev}{ev}
\DeclareMathOperator{\Fltr}{Fltr}
\DeclareMathOperator{\OPT}{OPT}

\newsavebox{\problemcontent}
\newenvironment{problembox}[1]{\par\smallskip
  \setlength{\fboxsep}{4pt}\begin{lrbox}{\problemcontent}\begin{minipage}{\dimexpr\linewidth-2\fboxsep-2\fboxrule\relax}\noindent\textsc{#1}.\space\ignorespaces
}{\end{minipage}\end{lrbox}\noindent\fbox{\usebox{\problemcontent}}\par\smallskip
}
 
\begin{document}

\title{When Elements Fail: Weighted Fault-Tolerant Matroid Bases}
\author{
Madhumita Kundu\inst{1} \and
Soumi Nandi\inst{2} \and
Saket Saurabh\inst{2,3}
}

\authorrunning{M. Kundu, S. Nandi and S. Saurabh}

\institute{
Ben-Gurion University, Be'er Sheva, Israel
\and
The Institute of Mathematical Sciences, Chennai, India
\and
University of Bergen, Norway
}
\hypersetup{
  pdftitle={When Elements Fail: Weighted Fault-Tolerant Matroid Bases},
  pdfauthor={},
  pdfsubject={Anonymous IPCO 2027 full version}
}
\maketitle

\begin{abstract}
Given a rank-\(r\) matroid on \(n\) elements with nonnegative costs and a partition into
safe and vulnerable elements, we seek a minimum-cost set that spans after every deletion
of at most \(k\) selected vulnerable elements. We resolve several open questions posed
by Bentert, Fomin, Golovach, and Morelle (ESA 2025) concerning weighted costs,
nonuniform failures, and explicitly specified failure scenarios.

To this end, we develop two compression methods. For matroids given by an independence
oracle, weighted flat-avoidance representatives yield a bounded subset of the ground set
containing an optimal solution and a \(2^{\mathcal{O}((r+k)^4)}n^{\mathcal{O}(1)}\)-time
algorithm. This establishes fixed-parameter tractability for weighted costs and nonuniform
failures simultaneously. For explicitly represented linear matroids, polynomial
certificates allow us to compress families of partial solutions while preserving
successful completions under all allowed failures. This yields a deterministic algorithm
using \(2^{\mathcal{O}(kr(r+k)\log(r+1))}n^{\mathcal{O}(1)}\) field operations over arbitrary
fields. Over fixed finite fields, a simpler reduction uses parallel classes.

For explicitly listed failure scenarios, we obtain fixed-parameter algorithms parameterized
by rank and the number of scenarios. In contrast, the size-budget decision version is W[1]-hard parameterized by rank
and maximum scenario size, even on uniform matroids with two-element scenarios.
\end{abstract}

\section{Introduction}\label{sec:introduction}

\par
Fault tolerance has been extensively studied in network design, matching, and
distributed storage. In these settings, the aim is to preserve a required function
despite the failure of some components---for example, maintaining connectivity after
link failures, retaining a perfect matching after edge deletions, or recovering data
when storage nodes become unavailable; see, e.g.,~\cite{cj,dimakis,hms}. A central
algorithmic question is how to provide the necessary redundancy at minimum cost.

For spanning problems, the need for redundancy is immediate. A basis spans a vector
space, but losing even one basis vector reduces its span. To withstand failures, we
must select a larger collection whose surviving vectors continue to span the original
space. Similarly, a spanning tree connects a graph, but additional edges are needed
to preserve connectivity after links fail. Matroids provide a common framework for
these two requirements through the notion of a spanning set.

Given a matroid \(M\) and a
nonnegative integer \(k\), the \textsc{Fault-Tolerant Basis} problem asks for a smallest
set \(B\subseteq E(M)\) such that \(B\setminus F\) spans \(M\) for every
\(F\subseteq B\) with \(|F|\le k\), or a report that no such set exists. For \(k=0\),
this is the ordinary basis problem; for \(k\ge1\), the selected set must contain enough
redundancy to withstand the allowed failures.

Bentert, Fomin, Golovach, and Morelle (ESA 2025)~\cite{bfgm} studied the parameterized
complexity of this problem. Their main result is an algorithm running in
\((kr)^{\mathcal{O}(kr^{4})}n^{\mathcal{O}(1)}\) time for rank-\(r\) matroids on \(n\)
elements given by an independence oracle, where \(k\) is the number of failures to be
tolerated. Thus, the problem is fixed-parameter tractable when parameterized by \(r+k\).

They left open the following questions:
\begin{itemize}
  \item \textbf{Weighted costs.} Does fixed-parameter tractability in \(r+k\)
  extend to the problem of finding a minimum-cost fault-tolerant spanning set
  when elements have arbitrary nonnegative costs?
  \item \textbf{Nonuniform failures.} Does this tractability extend to the model
  in which the ground set is partitioned into safe and vulnerable elements,
  and only vulnerable elements may fail?
  \item \textbf{Explicit failure scenarios.} Can their results be extended to
  the setting in which the possible failure scenarios are explicitly supplied
  as part of the input?
\end{itemize}

These questions are the starting point of our work. We first address weighted costs
and nonuniform failures together, and then study explicitly listed failure scenarios.
\par

\subsection{\texorpdfstring{Our Results and Tools}{Our Results and Tools}}\label{sec:compression}

\par
For ease of presentation, we first state the problem explicitly.
In \textsc{Weighted Nonuniform Fault-Tolerant Basis}, the input consists of a
rank-\(r\) matroid \(M=(E,\mathcal I)\), nonnegative integer costs
\(w:E\to\mathbb Z_{\ge0}\), a partition \(E=S\mathbin{\dot\cup}V\) into safe
and vulnerable elements, and a failure budget \(k\ge0\). The task is to find a
minimum-cost set \(B\subseteq E\) satisfying
\[
  \rank_M(B\setminus F)=r
  \qquad\text{for every }F\subseteq B\cap V\text{ with }|F|\le k,
\]
or report that no such set exists. The ordinary weighted model is the special case
\(S=\emptyset\). The surviving selected elements must span the original matroid;
the target rank does not change when a failure occurs.
\par

\par
Throughout, \(n=|E|\) and costs are encoded in binary. The cases \(r=0\) and
\(k=0\) are polynomial-time solvable; the exponential bounds below concern
\(r,k\ge1\). Section~\ref{sec:preliminaries} specifies the computational conventions.
\par

\par
Our first result shows that weighted costs and nonuniform failures can be handled
simultaneously: the problem remains fixed-parameter tractable when parameterized by
the rank and the failure budget. This answers the first two questions of Bentert et al.
Moreover, the algorithm requires only an independence oracle.
\par

\begin{theorem}\label{thm:oracle-intro}
\textup{\textsc{Weighted Nonuniform Fault-Tolerant Basis}} on rank-\(r\) matroids given by
independence oracles is solvable in
\[
  2^{\mathcal{O}\bigl(b\log b+br\log\binom{r+k-1}{k}\bigr)}n^{\mathcal{O}(1)},\;\text{ where } b=(k+1)r.
\]
In particular, the problem is solvable in \(2^{\mathcal{O}((r+k)^{4})}n^{\mathcal{O}(1)}\) time.

\end{theorem}

\par
The algorithm constructs an \emph{optimum-preserving core}: a subset of the ground
set whose size depends only on \(r\) and \(k\) and which contains a minimum-cost
feasible solution, whenever one exists. Section~\ref{sec:oracle} gives the construction
and its analysis.
\par

\par
When the matroid is given by an explicit linear representation, we can use its
algebraic structure to obtain a second algorithm with a different dependence on
\(r\) and \(k\). This algorithm is deterministic and works over arbitrary fields.
Its main ingredient is a compression method for families of partial solutions.
\par

\begin{theorem}\label{thm:linear-intro}
On an explicitly represented linear matroid, \textup{\textsc{Weighted Nonuniform Fault-Tolerant
Basis}} is solvable deterministically in
\[
  2^{\mathcal{O}(kr(r+k)\log(r+1))}n^{\mathcal{O}(1)}
\]
field operations. Over standard exact representations of finite fields and over the rationals,
the algorithm also has fixed-parameter bit complexity.

\end{theorem}

\par
The algorithm processes elements one at a time, retaining a bounded family of
partial selections. Whenever a partial selection admits a feasible completion,
a retained selection of no greater cost must work with that same completion under
all allowed failures. Polynomial certificates enforce this guarantee; we explain
the idea in the overview below.
\par

Section~\ref{sec:linear} gives the certificate, the dynamic program, and its full complexity
analysis.

\par
Over a fixed finite field, a simpler reduction is possible. A rank-\(r\)
representation has a number of nonloop parallel classes bounded in terms of \(r\)
alone, and we retain only a bounded number of inexpensive elements from each class. This gives the
following running-time bound.
\par

\begin{corollary}\label{cor:finite-intro}
For every fixed finite field \(\mathbb{F}_{Q}\), \textup{\textsc{Weighted Nonuniform
Fault-Tolerant Basis}} is solvable in
\[
  2^{\mathcal{O}(kr^{2}+kr\log(k+2))}n^{\mathcal{O}(1)}.
\]
In particular, for every fixed failure budget \(k\), the running time is
\(2^{\mathcal{O}(r^{2})}n^{\mathcal{O}(1)}\).
\end{corollary}

\par
The oracle and algebraic algorithms offer different advantages. When \(k=r\), their
running-time exponents are \(\mathcal{O}(r^{4})\) and
\(\mathcal{O}(r^{3}\log(r+1))\), respectively. For fixed \(r\ge2\) and growing
\(k\), the oracle bound has exponent \(\mathcal{O}(k\log(k+1))\), whereas the
algebraic bound has exponent \(\mathcal{O}(k^{2})\). Thus, on a represented linear
matroid, we may use whichever algorithm gives the better bound.
\par

\par
\begin{figure}[!hb]
\centering
\begin{tikzpicture}[
  box/.style={draw, rounded corners=2pt, align=center, font=\small,
              inner xsep=3pt, inner ysep=4pt},
  flow/.style={-{Stealth[length=4pt]}, semithick},
  x=1cm, y=1cm
]
\node[box,text width=10.5cm] (bound) at (0,0) {
  \textbf{Small-solution lemma}\\
  Some optimum has at most \(b=(k+1)r\) elements.
};
\node[box,text width=4.85cm] (oracle) at (-2.77,-1.64) {
  \textbf{Oracle algorithm}\\
  Process elements by cost; retain a bounded core separately for each safety type.
};
\node[box,text width=4.85cm] (linear) at (2.77,-1.64) {
  \textbf{Linear algorithm}\\
  Fix final size and vulnerable count; start with the empty partial selection.
};
\node[box,text width=4.85cm] (enumerate) at (-2.77,-3.46) {
  Enumerate subsets of the core of size at most \(b\).
};
\node[box,text width=4.85cm] (successors) at (2.77,-3.46) {
  For each input element, generate skip/take successors; group by safe/vulnerable counts and rank.
};
\node[box,text width=4.85cm] (test) at (-2.77,-5.07) {
  For each candidate, test spanning after every allowed failure.
};
\node[box,text width=4.85cm] (compress) at (2.77,-5.07) {
  Build common-completion certificates; keep a cost-ordered basis of coefficient vectors.
};
\node[box,text width=10.5cm] (return) at (0,-6.68) {
  Return a cheapest feasible selection, or report infeasibility.\\
  For the linear algorithm, check completed selections and compare all target counts.
};
\draw[flow] (bound.south -| oracle.north) -- (oracle);
\draw[flow] (bound.south -| linear.north) -- (linear);
\draw[flow] (oracle) -- (enumerate);
\draw[flow] (enumerate) -- (test);
\draw[flow] (linear) -- (successors);
\draw[flow] (successors) -- (compress);
\draw[flow] (test) -- (return.north -| test.south);
\draw[flow] (compress) -- (return.north -| compress.south);
\draw[flow] (compress.east) -- ++(.45,0) |-
  node[pos=.25,right,rotate=90,anchor=south,font=\scriptsize] {next element}
  (successors.east);
\end{tikzpicture}
\caption{Two alternative algorithms for weighted safe/vulnerable
fault-tolerant spanning. The oracle core contains an optimum. Linear
compression preserves every successful future completion at no greater cost.
The right-hand loop runs until all input elements have been processed.}
\label{fig:algorithms}
\end{figure}
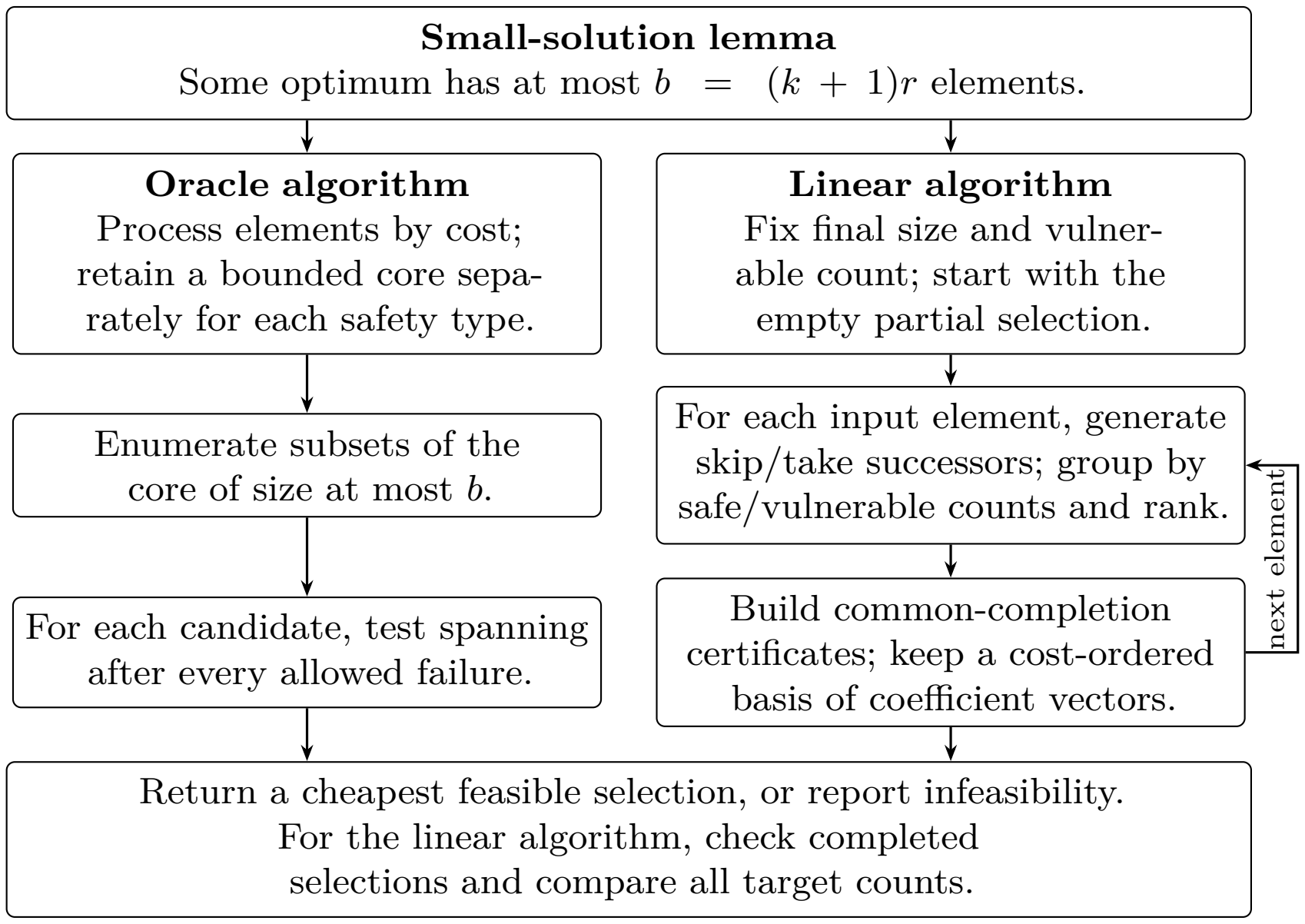

\smallskip\noindent\textbf{Overview of the compression methods.}
Both algorithms start from the observation that, whenever a feasible solution exists,
some minimum-cost solution contains at most \(b=(k+1)r\) elements. This follows by
extending the independent-layer argument of Bentert et al.~\cite{bfgm} to weighted
costs and nonuniform failures. However, enumerating all sets of at most \(b\) elements
takes \(n^{\mathcal{O}(b)}\) time, which does not give fixed-parameter tractability.
We therefore need to reduce the search space while preserving an optimum. Our two
algorithms do this in different ways: the oracle algorithm retains a bounded subset
of the ground set, whereas the linear algorithm retains bounded families of partial
selections. Figure~\ref{fig:algorithms} summarizes these two approaches.

\par

\par
\textit{An optimum-preserving oracle core:} We construct a retained set \(W\)
with the following replacement property. For every feasible selection \(B\) of size
at most \(b\) and every \(x\in B\setminus W\), there is an element
\(y\in W\setminus B\) of the same safety type as \(x\), with
\(w(y)\le w(x)\), such that \((B\setminus\{x\})\cup\{y\}\) remains feasible.
Here, having the same safety type means that both elements are safe or both are
vulnerable. Starting with an optimum of size at most \(b\), repeated replacements
therefore produce an optimum contained in \(W\). The challenge is to construct a
bounded set \(W\) that guarantees these replacements without knowing the optimum,
while preserving spanning under every allowed failure.

To understand which replacements are valid, fix a feasible selection \(B\) and an
element \(x\in B\), and write \(Y=B\setminus\{x\}\). Consider a failure set
\(F\subseteq Y\cap V\) with \(|F|\le k\). If \(Y\setminus F\) already spans,
no replacement is needed for this failure. Otherwise, feasibility of \(B\) implies
that adding \(x\) restores full rank, so \(Y\setminus F\) has rank \(r-1\).
Its closure is therefore a hyperplane, and a replacement \(y\) restores spanning
exactly when it lies outside that hyperplane. Thus, one element outside all these
hyperplanes works for every failure in which \(y\) survives. If \(y\) itself fails,
feasibility follows from the corresponding failure of \(x\) in \(B\).

Although there may be many failure sets, they define only a bounded number of
distinct obstructing hyperplanes. Indeed, the complement of each such hyperplane
in the restriction \(M|B\) is a cocircuit containing \(x\) and contained in
\(F\cup\{x\}\), so it has size at most \(k+1\). A deletion--contraction argument
bounds the number of these cocircuits by \(\binom{r+k-1}{k}\). Each obstructing hyperplane is the closure of a subset of \(B\), so its
intersection with \(B\) determines it; distinct hyperplanes therefore give distinct
cocircuits in \(M|B\).

This reduces the replacement task to avoiding a bounded family of hyperplanes.
Our weighted flat-avoidance construction processes elements in nondecreasing cost
order and limits the number retained in each flat according to its rank. It ensures
that every discarded element has a retained substitute of no greater cost outside
all the obstructing hyperplanes and outside the other selected elements. Applying
this construction separately to safe and vulnerable elements gives the replacement
property above, and hence the core used in Theorem~\ref{thm:oracle-intro}.

\textit{Polynomial certificates for common completions:} The linear algorithm
compresses families of partial selections with the same safe and vulnerable counts
and the same rank. For each partial selection \(P\), we construct a polynomial
\(G_P\). A future completion \(Z\) determines a symbolic substitution \(\ev_Z\),
common to all partial selections in the state, such that
\[
  \ev_Z(G_P)\not\equiv0
  \quad\Longleftrightarrow\quad
  P\cup Z\text{ survives every allowed failure}.
\]
For each failure pattern, a determinant tests whether the surviving past and future
elements span. We multiply these tests in one polynomial ring. After the substitution
for \(Z\), the product is nonzero exactly when every factor is nonzero, so one
certificate enforces all failure requirements for the same completion. Nonvanishing
here means that the resulting formal polynomial is not identically zero.

We then regard each \(G_P\) as a coefficient vector in a common monomial basis.
Processing partial selections in nondecreasing cost order, we retain a vector only
if it is linearly independent of the vectors already retained. Every discarded
polynomial is therefore a linear combination of retained polynomials of no greater
cost. Since the completion substitution is linear,
\[
  G_P=\sum_i\alpha_iG_{P_i},\qquad \ev_Z(G_P)\not\equiv0
  \quad\Longrightarrow\quad
  \ev_Z(G_{P_i})\not\equiv0\text{ for some }i.
\]
Thus, whenever a completion works for a discarded partial selection, it also works
for one retained selection of no greater cost. That one retained selection withstands
all allowed failures, because its certificate already combines all the failure tests.
This cost-ordered basis selection follows the representative-family paradigm~\cite{flps}.
The construction also relates to representative families under several matroid
constraints~\cite{vbtz}; here the failure requirements are combined in a shared
polynomial ring.

The main challenge is to construct the certificates in a coefficient space of bounded
dimension. We represent future columns through shared formal moment vectors and use
polynomial filters to remove failed future slots. This allows every determinant test
to use the same variables: taking their product increases the degree without introducing
new variables for each failure pattern. Bounding the number of variables and the degree
then bounds the number of retained partial selections. The construction uses formal
polynomials throughout and requires neither random evaluation nor a large field.
\par

\subsection{\texorpdfstring{Explicit Failure Scenarios}{Explicit Failure Scenarios}}\label{sec:scope}

\par
We next address the question of Bentert et al.~\cite{bfgm} concerning explicitly
specified failures. Instead of allowing every set of at most \(k\) vulnerable
elements to fail, the input supplies a family \(\mathcal F=\{F_1,\ldots,F_m\}\)
of failure scenarios. The task is to find a minimum-cost spanning set \(B\) such that
\(B\setminus F_i\) spans the matroid for every \(i\in[m]\). This is the
bulk-robust matroid basis model of Adjiashvili, Stiller, and Zenklusen~\cite{asz}.

Our results distinguish the number of scenarios from their maximum size.
Parameterized by the rank \(r\) and the number \(m\) of scenarios, the problem
is fixed-parameter tractable on independence-oracle matroids. For explicitly
represented linear matroids, we obtain an algorithm using
\(2^{\mathcal{O}(rm)}n^{\mathcal{O}(1)}\) field operations when \(m\ge1\).
In contrast, the size-budget decision version is W[1]-hard parameterized by rank
and maximum scenario size, even on uniform matroids when every scenario contains
only two elements. Thus, bounding the size of each failure scenario does not
provide the same tractability as bounding the number of scenarios.
Section~\ref{sec:scenarios} gives the algorithms and the hardness proof.
\par

\subsection{\texorpdfstring{Related Work}{Related Work}}\label{sec:related}

\par
The complexity on transversal matroids and gammoids was another question raised
by Bentert et al.~\cite{bfgm}. A construction of Adjiashvili, Bindewald, and
Michaels~\cite{abm} for robust assignment already implies NP-hardness on these
classes, even when only one element may fail. We present this construction in
matroid language and observe that its optimum identity also yields W[2]-hardness
parameterized by the solution budget minus the rank. Section~\ref{sec:gammoids}
gives the construction and the parameterized consequence in detail.

Adjiashvili, Stiller, and Zenklusen~\cite{asz} introduced the bulk-robust framework
and studied approximation algorithms for explicitly supplied failure scenarios,
including the matroid basis problem. Our scenario results concern exact algorithms
and their parameterized complexity. Robust-recoverable bases, as studied by
Ito et al.~\cite{ito}, impose overlap or limited-change requirements on a chosen
basis. In our model, the selected set purchases redundancy before any failure,
and the surviving elements must contain a basis after every allowed failure.

Dey and Kavitha~\cite{dey} study a related weighted preserver: after every bounded
failure, its surviving elements contain a minimum-cost basis of the surviving input
matroid. Their construction uses successive minimum-cost bases. The two objectives
charge for different sets. A basis preserver retains an optimal basis to be chosen
\emph{after} a failure; our objective minimizes the total cost of the set purchased
\emph{before} failures, subject to preserving the original rank. Consequently,
preserving a minimum-cost basis after each failure does not by itself ensure that
the retained ground set contains an optimum for our objective.
\par

\par
\textit{Organization.} Section~\ref{sec:preliminaries} gives the preliminaries.
Sections~\ref{sec:oracle} and~\ref{sec:linear} develop the oracle and linear-matroid
algorithms, respectively. Section~\ref{sec:scenarios} treats explicit failure scenarios,
and Section~\ref{sec:gammoids} presents the robust-assignment reduction and its
hardness consequences.
\par

\section{Preliminaries}\label{sec:preliminaries}

We use standard matroid terminology and parameterized-complexity notions. All ground sets and
failure families are finite. For an integer \(a\ge0\), write \([a]=\{1,\dots,a\}\), with
\([0]=\emptyset\). Throughout, \(n=|E|\) and costs are nonnegative integers given in binary. For
\(X\subseteq E\), write \(w(X)=\sum_{e\in X}w(e)\). A selected set need not be independent as
fault tolerance may require selecting more elements than the matroid rank.
We write \(L_{\mathrm{in}}\) for the total input encoding length; this is
distinct from the ground-set size \(n\).

\smallskip\noindent\textbf{Matroid terminology.} A matroid \(M=(E,\mathcal{I})\) consists of a ground set \(E\) and
a nonempty family \(\mathcal{I}\) of independent sets satisfying:

\begin{itemize}

\item \(\emptyset\in\mathcal{I}\),

\item if \(A\subseteq B\subseteq E\) and \(B\in\mathcal{I}\) then \(A\in\mathcal{I}\) and

\item if \(A,B\in\mathcal{I}\) with \(|A|<|B|\), some \(e\in B\setminus A\) satisfies
\(A\cup\{e\}\in\mathcal{I}\).

\end{itemize}

For \(X\subseteq E\), its rank is \(\rank_{M}(X)=\max\{|I|:I\subseteq X,I\in\mathcal{I}\}\).
Write \(r=\rank_{M}(E)\). A basis of \(X\) is a maximal independent subset of \(X\); it has size
\(\rank_{M}(X)\). A set \(X\) spans \(M\) if \(\rank_{M}(X)=r\). Adding or deleting one element
changes rank by at most one. The closure of \(X\) is
\[
  \cl_{M}(X)=\{e\in E:\rank_{M}(X\cup\{e\})=\rank_{M}(X)\}.
\]
We use the following standard closure properties:
\[
  X\subseteq\cl_M(X),\qquad
  \cl_M(\cl_M(X))=\cl_M(X),\qquad
  \rank_M(\cl_M(X))=\rank_M(X).
\]
Closure is monotone: if \(X\subseteq Y\), then \(\cl_M(X)\subseteq\cl_M(Y)\).
A flat is a set equal to its closure. For \(r\ge1\), a hyperplane is a flat of rank \(r-1\), and
a cocircuit is the complement of a hyperplane.
 A rank-zero matroid has no hyperplanes or cocircuits.

A circuit is an inclusion-minimal dependent set. An element \(e\) is a loop if \(\{e\}\) is
dependent and a coloop if it belongs to every basis. Two distinct nonloop elements are parallel
if they form a circuit. A matroid is simple if it has no loops or parallel pairs. We omit the
subscript \(M\) when the matroid is clear.

The restriction \(M|X\) has ground set \(X\) and independent sets \(\{I\in\mathcal{I}:I\subseteq
X\}\). Deletion is restriction to the remaining elements: \(M\setminus e=M|(E\setminus\{e\})\).
Contraction by \(e\), denoted \(M/e\), has rank function
\[
  \rank_{M/e}(X)=\rank_{M}(X\cup\{e\})-\rank_{M}(\{e\})\qquad (X\subseteq E\setminus\{e\}).
\]
For \(X\subseteq B\subseteq E\), closure in the restriction satisfies
\[
  \cl_{M|B}(X)=B\cap\cl_M(X).
\]
Indeed, restriction preserves the ranks of all subsets of \(B\), so the defining rank
test for closure is the same for each element of \(B\). The dual matroid \(M^*\) has
ground set \(E\) and has as its bases the complements of the bases of \(M\).
The uniform
matroid \(U_{r,E}\), where \(0\le r\le|E|\), has precisely the subsets of size at most \(r\) as
its independent sets. When only the ground-set size matters, we write \(U_{r,n}\).

\smallskip\noindent\textbf{Failure models and optimization problems.} In the safe/vulnerable model, the input
partitions \(E=S\mathbin{\dot\cup}V\) into safe and vulnerable elements and specifies a failure
budget \(k\ge0\). A selection \(B\subseteq E\) is \(k\)\textit{-fault tolerant} if
\begin{equation}
  \label{eq:failure-budget}
  \rank_{M}(B\setminus F)=r\qquad \text{for every}\; F\subseteq B\cap V\; \text{with}\; |F|\le k.
\end{equation}

The choice \(F=\emptyset\) includes the requirement that \(B\) itself spans. \textup{\textsc{Weighted Nonuniform Fault-Tolerant Basis}} asks for a minimum-cost such set, or
a report of infeasibility. \textit{\textup{\textsc{Ordinary Weighted Fault-Tolerant Basis}}} is
the special case \(S=\emptyset\), in which every selected element may fail. The explicit-scenario model, defined in Section~\ref{sec:scenarios}, instead specifies a list
of admissible failures. The corresponding unweighted size-budget decision
problems ask whether a feasible set of size at most a given integer \(\beta\) exists.
Here \(\beta\) is an input budget. We use \(b\) for a structural upper bound on the size
of some optimum, defined locally for each algorithm: \(b=(k+1)r\) in the failure-budget
model and \(b=mr\) in the algorithm for \(m\ge1\) listed scenarios.

\smallskip\noindent\textbf{Boundary cases.} If \(r=0\), the empty set is feasible and optimal in every model. For
the failure-budget models, \(k=0\) reduces to finding a minimum-cost basis by weighted matroid
greedy. The same algorithm handles the explicit-scenario model when \(m=0\). We may replace
\(k\) by \(\min\{k,|V|\}\) without changing feasibility. In the ordinary model, positive rank
and \(k\ge n\) imply infeasibility, since all selected elements may be deleted. Unless stated
otherwise, exponential parameter estimates for failure-budget algorithms concern \(r,k\ge1\);
polynomial boundary cases are treated separately. Loops may be deleted: removing selected loops
preserves every survivor rank and cannot increase cost.

\smallskip\noindent\textbf{Cocircuits as failure obstructions.} A set spans \(M\) exactly when it intersects every
cocircuit. Indeed, a nonspanning set is contained in a hyperplane and therefore misses its
complementary cocircuit; conversely, a set missing a cocircuit is contained in the complementary
hyperplane.
For rank zero, the condition is vacuous and every set spans.
This gives the following characterization.

\begin{proposition}\label{prop:cocircuit}
Let \(M\) be a rank-\(r\) matroid on \(E=S\mathbin{\dot\cup}V\), let \(k\ge0\) and let
\(B\subseteq E\). Then \(B\) is \(k\)-fault tolerant in the safe/vulnerable model if and only if
every cocircuit \(C^{*}\) of \(M\) satisfies
\[
  B\cap S\cap C^{*}\ne\emptyset\; \text{or}\; |B\cap V\cap C^{*}|\ge k+1.
\]
In particular, in the ordinary model this condition is \(|B\cap C^{*}|\ge k+1\) for every
cocircuit \(C^{*}\).

\end{proposition}

\begin{proof}

If a cocircuit contains no selected safe element and at most \(k\) selected vulnerable elements,
deleting those vulnerable elements leaves a set disjoint from that cocircuit, which does not
span. Conversely, under the displayed condition, every allowed failure leaves at least one
selected element in each cocircuit, so its survivors span. This also includes the no-failure
constraint.
\qed
\end{proof}

\smallskip\noindent\textbf{Linear representations and duality.} A linear representation over a field \(\mathbb{F}\)
assigns a column \(a_{e}\) to each \(e\in E\), with a set independent precisely when its columns
are linearly independent. Retaining \(r\) independent original rows gives an \(r\times n\)
representation of the same matroid. We use this form before applying exterior products or
determinant certificates. In a linear representation, parallel nonloop elements have columns
that are nonzero scalar multiples of one another.

The following characterization relates fault tolerance to independence in the dual.
It provides an alternative interpretation; the algorithms below do not require it.

\begin{proposition}\label{prop:parity-check}
Let \(M\) be a rank-\(r\) linear matroid over \(\mathbb{F}\), let \(B\subseteq E\) span \(M\)
and let \(A_{B}\) be an \(r\times|B|\) representation of \(M|B\) of row rank \(r\). Let
\(H_{B}\) have as its rows a basis of the orthogonal complement of the row space of \(A_{B}\) in
\(\mathbb{F}^{B}\). Thus \(H_B\) has \(|B|-r\) rows and \(|B|\) columns.
For \(F\subseteq B\), let \((H_{B})_{F}\) denote its columns indexed by
\(F\). Then
\[
  \rank_M(B\setminus F)<r\quad\Longleftrightarrow\quad (H_B)_F\text{ has linearly dependent columns}.
\]
Consequently, in the ordinary model \(B\) is \(k\)-fault tolerant exactly when every set of at
most \(k\) columns of \(H_{B}\) is independent. For \(k=2\), this is equivalent to simplicity of
\((M|B)^{*}\). In the safe/vulnerable model, only sets \(F\subseteq B\cap V\) of size at most
\(k\) need to be tested.

\end{proposition}

\begin{proof}

Deleting \(F\) lowers rank exactly when a nonzero linear combination of the rows of \(A_{B}\)
vanishes outside \(F\). Since \(\ker H_B\) is the row space of \(A_{B}\), this is equivalent to
a nonzero vector supported in \(F\) lying in \(\ker H_B\), or a linear dependence among the
columns of \((H_{B})_{F}\). The matrix \(H_{B}\) represents \((M|B)^{*}\), so independence of
all column sets of size at most two is exactly simplicity of that dual.
\qed
\end{proof}

\smallskip\noindent\textbf{Determinants and formal polynomial identities.} For matrices \(U\in\mathbb{F}^{a\times
m}\) and \(W\in\mathbb{F}^{m\times a}\), where \(0\le a\le m\), the Cauchy--Binet identity is
\begin{equation}
  \label{eq:cauchy-binet}
  \det(UW)=\sum_{\substack{L\subseteq[m]\\ |L|=a}}\det(U_{[:,L]})\det(W_{[L,:]}).
\end{equation}
Here \(U_{[:,L]}\) consists of the columns indexed by \(L\) and \(W_{[L,:]}\) consists of the
corresponding rows, both in increasing index order. If \(a>m\), the product has rank below \(a\)
and its determinant is zero, matching the empty sum. For indeterminates \(t_{1},\dots,t_{a}\),
the Vandermonde matrix \(V=(t^{j-1}_{i})^{a}_{i,j=1}\) satisfies
\[
  \det V=\prod_{1\le i<j\le a}(t_{j}-t_{i}).
\]
Empty products and determinants of \(0\times0\) matrices equal 1. These identities hold over
polynomial rings as well as over fields. A polynomial ring over a field is an integral domain,
so a product of polynomials is nonzero exactly when every factor is nonzero. Formal nonvanishing
is distinguished from nonvanishing after numerical evaluation, particularly over finite fields.

\smallskip\noindent\textbf{Computational conventions.} An independence oracle answers whether a queried set is
independent. Scanning a set \(X\) and retaining an element whenever it preserves independence
constructs a basis of \(X\), so a rank query uses at most \(|X|\) independence queries. Weighted
matroid greedy scans elements in nondecreasing cost order and retains those preserving
independence; it returns a minimum-cost basis. Ties are broken consistently. Oracle bounds count
independence queries and elementary arithmetic and comparison operations. With explicit
representations, we state field-operation bounds separately and use exact field arithmetic.
Bounds written with a factor \(n^{\mathcal{O}(1)}\) use this operation-count
convention; reading the encoded input and reducing an input matrix to \(r\) independent
rows contribute additional polynomial preprocessing. Bit bounds also account for the
encoding lengths of costs and field elements, and are stated in terms of
\(L_{\mathrm{in}}\), not only the number \(n\) of ground elements.
Weight additions and comparisons have polynomial bit cost in the encoded weights and the number
of selected elements. For standard finite-field encodings, field operations have polynomial bit
complexity in the field encoding length. Over \(\mathbb{Q}\), coefficient growth is analyzed in
the proof of the relevant algorithm. We make no bit-complexity claim for an unspecified
black-box field.

A parameterized problem is fixed-parameter tractable (FPT) with parameter \(t\) if it admits
a running time \(f(t)L_{\mathrm{in}}^{\mathcal{O}(1)}\), where \(f\) is computable and
\(L_{\mathrm{in}}\) is the input encoding length.
We use the standard notions of W[1]-hardness and W[2]-hardness under parameterized
reductions. Throughout, an optimum means a minimum-cost feasible selection; algorithms report
infeasibility when no feasible selection exists.

\section{A weighted FPT algorithm for oracle matroids}\label{sec:oracle}

We prove Theorem~\ref{thm:oracle-intro} by constructing a bounded set of original elements
that contains a minimum-cost feasible solution. Throughout this section, \(M\) is a rank-\(r\)
matroid given by an independence oracle, \(E=S\mathbin{\dot\cup}V\) is its partition into
safe and vulnerable elements, and \(w\) assigns nonnegative costs. A feasible selection must
span after every deletion of at most \(k\) of its vulnerable elements.
The \emph{safety type} of an element records whether it is safe or vulnerable.

The argument proceeds in five steps. First, some optimum has at most \(b=(k+1)r\) elements.
Second, we identify what is needed to replace one selected element: its replacement must have
the same safety type and avoid every hyperplane spanned by a deficient survivor set. Third,
we bound the number of these hyperplanes using small cocircuits through the removed element.
Fourth, a weight-ordered construction retains enough inexpensive elements to avoid any such
family of hyperplanes. Finally, we apply this construction to each safety type and search the
resulting core. The exchanges establish that the core contains an optimum; the algorithm
constructs the core without knowing that optimum.

For \(r=0\), the empty set is optimal, and for \(k=0\), weighted matroid greedy finds an
optimum. We therefore assume \(r,k\ge1\) in the algorithmic construction. Loops may be deleted
because they contribute nothing to any survivor rank and have nonnegative costs. The auxiliary
lemmas below state their own assumptions, including the boundary cases in which they apply.

\subsection{Small optimal solutions}\label{sec:small-solutions}

We begin by bounding the number of elements needed in an optimal solution. The following
sparsification lemma preserves the entire closure after every failure of size at most \(k\). We use a disjoint-layer construction, as in the dimension-reduction procedure for
\textup{\textsc{Matrix Rigidity}} \cite{flmssz} and the small-solution bound for
\textup{\textsc{Fault-Tolerant Basis}} \cite[full version, Proposition~3]{bfgm}.

\begin{lemma}\label{lem:sparsification}
Let \(M=(E,\mathcal{I})\) be a matroid of rank \(r\), let \(B\subseteq E\), and let \(k\ge0\) be
an integer. There exists a set \(C\subseteq B\) with \(|C|\le(k+1)r\) such that
\[
  \cl_{M}(C\setminus F)=\cl_{M}(B\setminus F)\qquad \text{for every}\; F\subseteq B\; \text{with}\; |F|\le k.
\]
Consequently, every feasible instance of \textup{\textsc{Weighted Nonuniform Fault-Tolerant
Basis}} has a minimum-cost feasible solution of size at most \(b=(k+1)r\). The same conclusion
holds when the allowed failures form a prescribed family of subsets of \(E\), each of size at
most \(k\).

\end{lemma}

\begin{proof}

 For each
\(j\in\{0,\dots,k\}\), choose a maximal independent subset \(I_{j}\) of
\(B\setminus\bigcup_{h<j}I_{h}\), and put \(C=\bigcup^{k}_{j=0}I_{j}\). The sets
\(I_{0},\dots,I_{k}\) are pairwise disjoint and each has at most \(r\) elements, so
\(|C|\le(k+1)r\).

Every element \(e\in B\setminus C\) was available when each \(I_{j}\) was chosen. By maximality
of \(I_{j}\), the set \(I_{j}\cup\{e\}\) is dependent, and hence \(e\in\cl_{M}(I_{j})\). Thus
every element of \(B\setminus C\) is spanned by each layer.

Fix \(F\subseteq B\) with \(|F|\le k\). Since the \(k+1\) layers are pairwise disjoint, at least
one layer \(I_{j}\) is disjoint from \(F\). Therefore \(I_{j}\subseteq C\setminus F\), and
\[
  (B\setminus C)\setminus F\subseteq\cl_{M}(I_{j})\subseteq\cl_{M}(C\setminus F).
\]
Every element of \(B\setminus F\) either belongs to \(C\setminus F\) or to \((B\setminus
C)\setminus F\). It follows that \(B\setminus F\subseteq \cl_{M}(C\setminus F)\). Taking
closures and using idempotence gives \(\cl_{M}(B\setminus F)\subseteq\cl_{M}(C\setminus F)\).
The reverse inclusion follows from \(C\setminus F\subseteq B\setminus F\), proving the equality.

Now let \(B\) be a minimum-cost feasible solution and apply the construction above. Every
failure \(F\subseteq C\cap V\) with \(|F|\le k\) is also an allowed failure for \(B\).
Feasibility of \(B\) and the closure equality imply \(\cl_{M}(C\setminus F)=\cl_{M}(B\setminus
F)=E\). Thus \(C\) is feasible. Since \(C\subseteq B\) and costs are nonnegative, \(w(C)\le
w(B)\), so \(C\) is also optimal.

Finally, suppose the allowed failures form a prescribed family of subsets of \(E\), each of size
at most \(k\). For any such failure \(F\), apply the closure equality to \(F\cap B\). Since
\(C\subseteq B\), we have \(C\setminus(F\cap B)=C\setminus F\) and \(B\setminus(F\cap
B)=B\setminus F\). Hence feasibility is preserved under every prescribed failure. The same cost
argument proves the final assertion.
\qed
\end{proof}

\begin{remark}\label{rem:size-tightness}

The size bound is tight even for ordinary unweighted fault tolerance. For example, take the
direct sum of \(r\) rank-one uniform matroids, each with exactly \(k+1\) elements, and make
every element vulnerable. A feasible set must select all \(k+1\) elements of every summand. Otherwise, an allowed failure can delete all selected elements in that summand. Thus its
unique feasible set has size \((k+1)r\).

\end{remark}

\subsection{Replacing a selected element}\label{sec:oracle-replacement}

The size bound does not yet give an FPT algorithm: searching all subsets of \(E\) of size at
most \(b\) takes \(n^{\mathcal{O}(b)}\) time. We seek a bounded set \(W\subseteq E\) into which
an optimum can be moved by exchanges. To determine what \(W\) must preserve, we first study a
single exchange while fixing the other selected elements.

\begin{definition}\label{def:replacement-equivalence}
Fix \(Y\subseteq E\). Two elements \(x,y\in E\setminus Y\) are equivalent relative to \(Y\)
if they have the same safety type---either both belong to \(S\) or both belong to \(V\)---and
\[
  \rank_M((Y\setminus F)\cup\{x\})
  =\rank_M((Y\setminus F)\cup\{y\})
\]
for every \(F\subseteq Y\cap V\) with \(|F|\le k\).
\end{definition}

For fixed \(Y\), this is an equivalence relation: it compares the safety type and the ranks
obtained after each allowed failure within \(Y\). Costs are not part of the relation. In an
exchange, we will additionally require \(w(y)\le w(x)\).

Now fix a feasible set \(B\), choose \(x\in B\), and put \(Y=B\setminus\{x\}\). Define
\[
  \mathcal H(B,x)=
  \bigl\{\cl_M(Y\setminus F): F\subseteq Y\cap V,\ |F|\le k,
               \ \rank_M(Y\setminus F)<r\bigr\}.
\]
We regard this as a set of distinct flats, so different failures may contribute the same member.
Feasibility of \(B\) implies that adding \(x\) to each survivor set restores rank \(r\).
Thus every member of \(\mathcal H(B,x)\) is a hyperplane avoiding \(x\).

\begin{lemma}\label{lem:replacement-criterion}
Let \(B\) be feasible, let \(x\in B\), and put \(Y=B\setminus\{x\}\). An element
\(y\in E\setminus B\) of the same safety type as \(x\) is equivalent to \(x\) relative to
\(Y\) if and only if it avoids every member of \(\mathcal H(B,x)\). In this case,
\(B'=Y\cup\{y\}\) is feasible. If also \(w(y)\le w(x)\), then \(w(B')\le w(B)\).
\end{lemma}

\begin{proof}
Fix \(F\subseteq Y\cap V\) with \(|F|\le k\). Since \(B\) is feasible,
\(\rank_M((Y\setminus F)\cup\{x\})=r\). Deleting one element lowers rank by at most one,
so \(Y\setminus F\) has rank either \(r\) or \(r-1\). If it has rank \(r\), adding either
\(x\) or \(y\) leaves the rank unchanged. Otherwise, \(H_F=\cl_M(Y\setminus F)\) is a
hyperplane avoiding \(x\), and adding \(y\) restores rank \(r\) exactly when \(y\notin H_F\).
Applying this observation to every allowed \(F\) proves the equivalence.

To verify feasibility of \(B'\), let \(F'\subseteq B'\cap V\) have size at most \(k\).
If \(y\notin F'\), then \(F'\subseteq Y\cap V\). Equivalence gives
\[
  \rank_M(B'\setminus F')
  =\rank_M((Y\setminus F')\cup\{y\})
  =\rank_M((Y\setminus F')\cup\{x\})=r.
\]
If \(y\in F'\), then \(y\) is vulnerable, and hence so is \(x\). The set
\[
  F=(F'\setminus\{y\})\cup\{x\}
\]
is an allowed failure for \(B\), with \(|F|=|F'|\) and
\(B'\setminus F'=B\setminus F\). Feasibility of \(B\) again gives rank \(r\).
Finally, \(w(B')=w(B)-w(x)+w(y)\le w(B)\) whenever \(w(y)\le w(x)\).
\qed
\end{proof}

The same replacement must avoid all members of \(\mathcal H(B,x)\). A separate replacement
for each failure would not yield one feasible selected set. We next bound the size of this
family, and then construct representatives that work for every family of that size.

\subsection{Bounding the replacement obstructions}\label{sec:cocircuit-bound}

Each deficient survivor hyperplane gives a small cocircuit through \(x\) in the restriction
\(M|B\). We first bound such cocircuits in an arbitrary matroid, and then verify that this
also bounds the distinct ambient hyperplanes in \(\mathcal H(B,x)\).

\begin{lemma}\label{lem:rooted-cocircuits}
Let \(M\) be a matroid of rank \(r\ge1\), let \(x\in E(M)\), and let
\(k\ge0\) be an integer. Then the number of cocircuits
\(C^{*}\) satisfying
\[
  x\in C^{*},\qquad |C^{*}|\le k+1
\]
is at most
\[
  \binom{r+k-1}{k}.
\]
\end{lemma}

\begin{proof}

It is equivalent to count the hyperplanes \(H\) avoiding \(x\) whose complements have size at
most \(k+1\). We prove the bound by induction on \(r+k\). If \(x\) is a loop, the count is zero.
So now we assume that \(x\) is not a loop. For \(r=1\) there is a unique hyperplane, namely the
set of loops. For \(k=0\), the only possible cocircuit is \(\{x\}\). These two cases also
satisfy the bound.

Assume now that \(r\ge2\) and \(k\ge1\). If \(x\) is a coloop, the only cocircuit
containing it is \(\{x\}\), so the bound holds. Otherwise, \(x\) is neither a loop nor a
coloop and belongs to a circuit \(C\) of size at least two. Choose
\(y\in C\setminus\{x\}\). Since \(y\) belongs to this circuit, it is neither a loop nor
a coloop. We separate the counted hyperplanes according to whether they contain \(y\).

If \(y\in H\), then \(H\setminus\{y\}\) is a flat of \(M/y\). Since \(y\) is a nonloop, the flat
has rank \(r-2\) and \(M/y\) has rank \(r-1\). Thus it is a hyperplane with complement
\(E(M)\setminus H\), which contains \(x\) and has size at most \(k+1\). Adjoining \(y\) recovers
\(H\), so the correspondence is injective. Induction bounds this case by \(\tbinom{r+k-2}{k}\).

If \(y\notin H\), then \(H\) remains a flat of rank \(r-1\) in \(M\setminus y\). Since \(y\) is
not a coloop, this deletion has rank \(r\), and \(H\) is still a hyperplane. Its complement is
\((E(M)\setminus H)\setminus\{y\}\), which contains \(x\) and has size at most \(k\). Induction
bounds the number of these hyperplanes by \(\tbinom{r-1+k-1}{k-1}=\tbinom{r+k-2}{k-1}\). Summing
the two bounds and applying Pascal's identity completes the induction.

The bound is attained by \(U_{r,r+k}\). Every cocircuit of this uniform matroid has size
\(k+1\), and exactly \(\tbinom{r+k-1}{k}\) of them contain a prescribed element.
\qed
\end{proof}

\begin{lemma}\label{lem:deficient-hyperplanes}
For every feasible set \(B\) and every \(x\in B\),
\[
  |\mathcal H(B,x)|\le \binom{r+k-1}{k}.
\]
\end{lemma}

\begin{proof}
Put \(Y=B\setminus\{x\}\), and consider a failure \(F\subseteq Y\cap V\) with \(|F|\le k\)
that contributes \(H_F=\cl_M(Y\setminus F)\) to \(\mathcal H(B,x)\). As observed above,
\(\rank_M(Y\setminus F)=r-1\) and \(x\notin H_F\).

The restriction \(M|B\) has rank \(r\), since \(B\) is feasible. The set \(B\cap H_F\)
is a flat of this restriction. It contains \(Y\setminus F\) and is contained in \(H_F\),
so its rank is \(r-1\). Consequently,
\[
  C_F=B\setminus H_F
\]
is a cocircuit of \(M|B\) containing \(x\). Every element of \(Y\setminus F\) belongs to
\(H_F\), so \(C_F\subseteq F\cup\{x\}\) and \(|C_F|\le k+1\).

Distinct ambient hyperplanes yield distinct cocircuits. Indeed,
\[
  \cl_M(B\cap H_F)=H_F,
\]
because \(B\cap H_F\) contains \(Y\setminus F\), whose closure is \(H_F\), and is itself
contained in the flat \(H_F\). Equal cocircuits give equal complements within \(B\), hence
equal closures in \(M\), and therefore equal hyperplanes. Lemma~\ref{lem:rooted-cocircuits}
applied to \(M|B\) now proves the bound.
\qed
\end{proof}

We therefore need representatives that avoid any family of at most
\(t=\binom{r+k-1}{k}\) hyperplanes avoiding a discarded element. We also need a reserve of
\(b\) inexpensive candidates: at most \(b-1\) of them may already belong to the selection
\(Y\). The next lemma provides exactly this guarantee without knowing \(B\), \(x\), or its
family of deficient hyperplanes in advance.

\subsection{Weighted flat-avoidance representatives}\label{sec:saturation}

We now construct representatives for the requirement identified above. The lemma is stated
for arbitrary positive integers \(b,t\), so it can also be used for the explicit-scenario
model later. It processes a set \(A\) in nondecreasing cost order and limits the number of
retained elements in each flat according to its rank.

The lemma selects a small set \(W\subseteq A\) of representatives. For every discarded element
\(x\in A\setminus W\) and every collection of at most \(t\) hyperplanes not containing \(x\), at
least \(b\) elements of \(W\) have cost at most \(w(x)\) and lie outside all these hyperplanes.
For example, consider a rank-two matroid represented by nonzero vectors in \(\mathbb{R}^{2}\),
and let \(t=2\) and \(b=3\). Choose any two lines through the origin, neither containing the
discarded vector \(x\). The lemma guarantees that \(W\) contains at least three vectors outside
both lines, each of cost at most \(w(x)\). Even if two of these vectors are already selected, a
third remains available. Like \(x\), this vector can extend a rank-one set lying on either line
to a set of rank two.

\begin{lemma}\label{lem:flat-avoidance}
Let \(M\) be a loopless matroid of rank \(r\ge1\), let \(A\subseteq E(M)\), let
\(w:A\to\mathbb{R}_{\ge0}\), and let \(b,t\ge1\) be integers. Define \(L_{0}=0\) and
\[
  L_{h}=tL_{h-1}+b=b\sum^{h-1}_{j=0}t^{j}\qquad \text{for}\; 1\le h\le r.
\]
There is an algorithm that computes a set \(W\subseteq A\) of size at most \(L_{r}\) with the
following property. For every \(x\in A\setminus W\), every collection of at most \(t\)
hyperplanes \(H_{1},\dots,H_{\ell}\) with \(x\notin H_{i}\) for each \(i\), and every set
\(D\subseteq E(M)\) of size at most \(b-1\), there exists
\[
  y\in W\setminus\Bigl(D\cup\bigcup^{\ell}_{i=1}H_{i}\Bigr)\qquad \text{such that}\qquad w(y)\le w(x).
\]
The algorithm uses \(\mathcal{O}(nr(L_{r}+1)^{r+1})\) independence queries, in addition to
sorting the elements of \(A\) by weight, where \(n=|E(M)|\).

\end{lemma}

\begin{proof}

First we process the elements of \(A\) in nondecreasing
order of cost, breaking ties arbitrarily. Initially we set \(W=\emptyset\). Throughout the
process we maintain
\begin{equation}
  \label{eq:flat-invariant}
  |W\cap Q|\le L_{\rank(Q)}\qquad \text{for every flat }Q\text{ of }M.
\end{equation}
We call a flat \(Q\) \emph{saturated} if \(|W\cap Q|=L_{\rank(Q)}\).

Now when an element \(x\in A\) is considered, we discard \(x\) if there exists a saturated
flat \(Q\) containing \(x\). Otherwise, we add \(x\) to \(W\). We verify that
\eqref{eq:flat-invariant} is preserved. An insertion of \(x\) can change the quantity \(|W\cap
Q|\) only for flats \(Q\) containing \(x\). Before the insertion, every such flat satisfies
\(|W\cap Q|<L_{\rank(Q)}\), for otherwise, it would be saturated and \(x\) would have been
discarded. Since the quantities involved are integral, after inserting \(x\) we still have
\(|W\cap Q|\le L_{\rank(Q)}\). Thus the invariant is maintained.

Observe that applying \eqref{eq:flat-invariant} to the flat \(E(M)\), which has rank \(r\),
gives \(|W|\le L_{r}\), throughout the algorithm.

We next show that the above rule can be implemented
without enumerating all flats of \(M\).

First observe that the thresholds are strictly increasing: \(L_{h}=tL_{h-1}+b>L_{h-1}\), and
hence \(L_{h}>L_{d}\) whenever \(h>d\). Suppose that \(Q\) is a saturated flat of rank \(h\). We
claim that \(\rank(W\cap Q)=h\). Assume instead that \(d:=\rank(W\cap Q)<h\), and define
\(P:=\cl(W\cap Q)\). Then \(P\) is a flat of rank \(d\). Moreover, since \(Q\) is a flat
containing \(W\cap Q\), closure minimality gives \(P=\cl(W\cap Q)\subseteq Q\). On the other
hand, \(W\cap Q\subseteq\cl(W\cap Q)=P\). Combining this with \(P\subseteq Q\) yields \(W\cap
P=W\cap Q\). Indeed, the inclusion \(P\subseteq Q\) gives \(W\cap P\subseteq W\cap Q\), whereas
\(W\cap Q\subseteq P\) gives the reverse inclusion.

Since \(Q\) is saturated, \(|W\cap P|=|W\cap Q|=L_{h}\). But \(\rank(P)=d<h\), and therefore
\(L_{h}>L_{d}\). Consequently
\[
  |W\cap P|=L_{h}>L_{d}=L_{\rank(P)},
\]
contradicting the invariant \eqref{eq:flat-invariant}. Hence \(\rank(W\cap Q)=h\). It follows
that \(W\cap Q\) contains an independent set \(I\) of size \(h\) spanning \(Q\). Thus
\(\cl(I)=Q\). Therefore every saturated flat has the form \(\cl(I)\) for some independent set
\(I\subseteq W,|I|\le r\).

Accordingly, when processing \(x\), enumerate all independent subsets \(I\subseteq W,|I|\le r\).
For each such \(I\), test whether \(x\in\cl(I)\) and determine the quantity \(|W\cap\cl(I)|\).
Discard \(x\) precisely when some enumerated \(I\) satisfies
\[
  x\in\cl(I),|W\cap\cl(I)|=L_{|I|}.
\]
Every such witness indeed defines a saturated flat, because \(I\) is independent and hence
\(\rank(\cl(I))=|I|\). Conversely, if \(x\) lies in a saturated flat \(Q\), the preceding
argument produces an independent set \(I\subseteq W\cap Q\) of size \(\rank(Q)\) with
\(\cl(I)=Q\), and this \(I\) is among the enumerated subsets. Hence the finite procedure is
exactly equivalent to the abstract saturation rule.

Now for an independent set \(I\), membership of an element \(e\) in \(\cl(I)\) can be tested by
checking whether \(I\cup\{e\}\) is independent: for \(e\notin I\),
\[
  e\in\cl(I)\quad\Longleftrightarrow\quad I\cup\{e\}\text{ is dependent},
\]
while membership is automatic when \(e\in I\).

Since \(|W|\le L_{r}\), the number of subsets of \(W\) of size at most \(r\) is at most
\[
  \sum_{j=0}^r\binom{|W|}{j}\le\sum_{j=0}^r\binom{L_r}{j}.
\]
Using \(\tbinom{L_r}{j}\le L_r^j\), we obtain
\[
  \sum_{j=0}^r\binom{L_r}{j}\le\sum_{j=0}^r L_r^j\le(L_r+1)^r.
\]
Testing whether an enumerated set \(I\) is independent uses one query.
For each independent \(I\), testing \(x\in\cl(I)\) and determining
\(W\cap\cl(I)\) uses at most \(\mathcal{O}(L_r+1)\) further independence queries. Thus one may in fact implement the rule
using \(\mathcal{O}\bigl(n(L_{r}+1)^{r+1}\bigr)\) independence queries. In particular, this is
within the stated bound \(\mathcal{O}\bigl(nr(L_{r}+1)^{r+1}\bigr)\).

To prove the replacement guarantee, fix a discarded element \(x\), and let \(Q\) be a saturated
witness at the moment when \(x\) is discarded. Write \(h:=\rank(Q)\). Let \(Z:=W\cap Q\) at that
moment. Since \(Q\) is saturated, \(|Z|=L_{h}\). All elements of \(Z\) were processed before
\(x\), and therefore \(w(z)\le w(x)\) for every \(z\in Z\). Since elements are never removed
from \(W\), all elements of \(Z\) remain in the final set \(W\).

Because \(M\) is loopless and \(x\in Q\), we have \(h\ge1\). Now let \(H\) be any hyperplane
avoiding \(x\). Since \(x\in Q\) but \(x\notin H\), we have \(Q\nsubseteq H\). Therefore \(Q\cap
H\) is a proper flat of \(Q\), and hence \(\rank(Q\cap H)\le h-1\). Applying the invariant
\eqref{eq:flat-invariant} to the flat \(Q\cap H\) gives
\[
  |W\cap(Q\cap H)|\le L_{\rank(Q\cap H)}\le L_{h-1}.
\]
Since \(Z\subseteq W\cap Q\), it follows that \(|Z\cap H|\le L_{h-1}\).

Now let \(H_{1},\dots,H_{\ell}\) be at most \(t\) hyperplanes avoiding \(x\). By the union
bound,
\[
  \biggl|Z\cap\bigcup_{i=1}^{\ell}H_i\biggr|\le\sum_{i=1}^{\ell}|Z\cap H_i|\le tL_{h-1}.
\]
Hence
\[
  \biggl|Z\setminus\bigcup_{i=1}^{\ell}H_i\biggr|\ge L_h-tL_{h-1}=b.
\]
Finally, since \(|D|\le b-1\), at least one element of \(Z\setminus\bigcup^{\ell}_{i=1}H_{i}\)
lies outside \(D\). Thus there exists
\[
  y\in Z\setminus\Bigl(D\cup\bigcup^{\ell}_{i=1}H_{i}\Bigr).
\]
Because \(Z\subseteq W\) and every element of \(Z\) has cost at most \(w(x)\), this \(y\)
satisfies
\[
  y\in W\setminus\Bigl(D\cup\bigcup^{\ell}_{i=1}H_{i}\Bigr),w(y)\le w(x),
\]
as required.
\qed
\end{proof}

\subsection{Constructing and searching the core}\label{sec:oracle-core}

We apply the representative construction separately to safe and vulnerable elements.
The resulting union contains an optimum by the replacement criterion, and its bounded size
permits exhaustive search.

\begin{theorem}\label{thm:oracle}
On rank-\(r\) matroids given by independence oracles,
\textup{\textsc{Weighted Nonuniform Fault-Tolerant Basis}}
is fixed-parameter tractable in \(r+k\). The cases \(r=0\) and
\(k=0\) are polynomial-time solvable. For \(r,k\ge1\), the running time is
\[
  2^{\mathcal{O}\bigl(b\log b+br\log\binom{r+k-1}{k}\bigr)}n^{\mathcal{O}(1)},\qquad b=(k+1)r.
\]
In particular, the running time is \(2^{\mathcal{O}((r+k)^{4})}n^{\mathcal{O}(1)}\).

\end{theorem}

\begin{proof}

For \(r=0\), return the empty set. For \(k=0\), return a minimum-cost basis using the matroid
greedy algorithm. Henceforth assume \(r,k\ge1\).

To construct the core, delete all loops, which cannot improve a nonnegative-cost solution.
Set
\[
  t=\binom{r+k-1}{k},\qquad b=(k+1)r.
\]
Apply Lemma~\ref{lem:flat-avoidance} separately to the safe set \(S\) and the vulnerable set
\(V\), with these values of \(b\) and \(t\). The flat tests in both applications use the ambient
matroid \(M\). Let \(W\) be the union of the two resulting cores. Then
\begin{equation}
  \label{eq:core-size}
  |W|\le2b\sum^{r-1}_{j=0}t^{j}.
\end{equation}
The factor two is unnecessary when all elements are vulnerable.

To prove that \(W\) contains an optimum, use Lemma~\ref{lem:sparsification} to choose a
minimum-cost feasible set \(B\) with \(|B|\le b\). Suppose \(x\in B\setminus W\), and put
\(Y=B\setminus\{x\}\). Lemma~\ref{lem:deficient-hyperplanes} gives at most \(t\) members
in \(\mathcal H(B,x)\), each avoiding \(x\). Apply Lemma~\ref{lem:flat-avoidance} in the
subset \(S\) or \(V\) containing \(x\), with forbidden set \(D=Y\). Since \(|Y|\le b-1\), it gives
an element \(y\in W\setminus Y\) of the same safety type, with \(w(y)\le w(x)\), outside
every member of \(\mathcal H(B,x)\). Also \(y\ne x\), because \(x\notin W\), so
\(y\notin B\). Lemma~\ref{lem:replacement-criterion} shows that \(B'=Y\cup\{y\}\) is
feasible and has no greater cost.

We have \(|B'|=|B|,w(B')\le w(B)\), and \(|B'\setminus W|=|B\setminus W|-1\). Since \(B\) was
optimal, \(B'\) is optimal as well. Repeating this exchange at most \(b\) times produces an
optimum contained in \(W\). This is an existence argument; the algorithm need not identify an
optimum or enumerate its deficient hyperplanes when constructing \(W\).

To search the core, enumerate all subsets of \(W\) of size at most \(b\). For each
candidate, test every subset of its vulnerable elements of size at most \(k\) as a failure,
using greedy rank computations on the survivors. Return a cheapest feasible candidate, or report
infeasibility if none exists. The exchange argument proves correctness in both cases.

Equation \eqref{eq:core-size} and \(r\le b\) imply
\[
  \log(|W|+1)=\mathcal{O}(\log b+r\log t).
\]
There are at most \((|W|+1)^{b+1}\) candidates. Each has at most \(2^{b}\) failure subsets, and
each survivor rank can be computed with at most \(b\) independence queries. Since \(b\ge2\),
these verification costs are absorbed in
\[
  2^{\mathcal{O}(b\log b+br\log t)}n^{\mathcal{O}(1)}.
\]
For core construction, the query bound of Lemma~\ref{lem:flat-avoidance} has parameter-dependent
logarithm
\[
  \mathcal{O}\bigl((r+1)(\log b+r\log t)\bigr).
\]
This is within the same bound because \(r+1\le b\). Cost sorting and arithmetic contribute
polynomial overhead under the computational conventions of Section~\ref{sec:preliminaries}.
Finally, \(b\le (r+k+1)^{2}\) and \(\log t\le r+k\), yielding the stated quartic-exponent upper
bound.
\qed
\end{proof}

\begin{remark}\label{rem:oracle-kernel}

The core in Theorem~\ref{thm:oracle} is a subset of the original ground set and contains a
minimum-cost feasible solution whenever one exists. In the pure independence-oracle model, this
does not by itself provide a polynomial-bit kernel: encoding the restricted matroid as a
standalone instance is a separate issue.

\end{remark}

\section{Linear representations}\label{sec:linear}

Section~\ref{sec:oracle} constructs a bounded subset of the ground set that contains an
optimum. With an explicit linear representation, we can instead process the elements one at
a time and maintain a bounded family of partial selections. Taking or skipping the next
element generates new candidates; compression then retains a small subfamily. The main
question is what this subfamily must preserve so that an optimum remains reachable.

Throughout this section, \(M\) is represented over a field \(\mathbb F\), its ground set is
partitioned into safe elements \(S\) and vulnerable elements \(V\), and costs are
nonnegative. Retaining a basis of the row space lets us assume that the representing columns
belong to \(\mathbb F^r\), where \(r=\rank(M)\). For rank zero, the empty set is optimal;
for \(k=0\), a minimum-cost basis is optimal. The parameter estimates for the fault-tolerant
algorithm therefore assume \(r,k\ge1\).

\subsection{What compression must preserve}\label{sec:completion-property}

Fix an order \(e_1,\dots,e_n\) of the ground set and a processed prefix
\(E_j=\{e_1,\dots,e_j\}\). A \emph{partial selection} is a set \(P\subseteq E_j\), and a
\emph{future completion} is a set \(Z\subseteq E\setminus E_j\). We call \(Z\) successful
for \(P\) if \(P\cup Z\) is feasible, that is,
\[
  \rank_M\bigl((P\cup Z)\setminus F\bigr)=r
  \quad\text{for every }F\subseteq(P\cup Z)\cap V\text{ with }|F|\le k.
\]
The algorithm will group partial selections by their numbers of safe and vulnerable
elements. Replacing a selection within such a group preserves these counts, so the same
completion still gives the same final counts.

\begin{definition}
\label{def:common-completion-representatives}
Let \(\mathcal P\) be a family of subsets of \(E_j\), all with the same numbers of safe and
vulnerable elements. A subfamily \(\widehat{\mathcal P}\subseteq\mathcal P\) is
\emph{representative for common completions} if, for every \(P\in\mathcal P\) and every
\(Z\subseteq E\setminus E_j\) successful for \(P\), there exists
\(P'\in\widehat{\mathcal P}\) such that
\[
  w(P')\le w(P)
  \qquad\text{and}\qquad
  P'\cup Z\text{ is feasible}.
\]
When final safe and vulnerable counts are prescribed, we require this property only for
completions \(Z\) meeting those counts.
\end{definition}

Here \(P'\) may depend on \(P\) and \(Z\), but it must be chosen before the failure occurs:
one retained partial selection must work with the fixed set \(Z\) under every allowed
failure of \(P'\cup Z\). Preserving a separate representative for each failure is therefore
insufficient. We also preserve the completion \(Z\) itself, rather than merely the existence
of some completion. At the final prefix, \(Z=\emptyset\), and the property ensures that an
optimum has a retained feasible replacement of no greater cost.

We obtain this guarantee from a cost-ordered linear-basis construction. First we state the
underlying linear-algebra lemma and illustrate how it preserves simultaneous spanning
constraints. We then encode all allowed failures in one polynomial: substituting a fixed
completion gives a nonzero polynomial exactly when that completion is successful. Applying
the same linear-basis construction to coefficient vectors yields the required representatives.
The resulting dynamic program is given in Theorem~\ref{thm:linear}. Finally,
Section~\ref{sec:finite-fields} gives a simpler ground-set reduction for fixed finite fields.

\subsection{Compression by linear representatives}\label{sec:linear-representatives}

To obtain the representatives of Definition~\ref{def:common-completion-representatives},
we will associate a vector \(v(P)\) with each partial selection \(P\) in a fixed group.
A future completion \(Z\) will determine a linear map \(L_Z\), common to the group, such that
\[
  L_Z(v(P))\ne0
  \quad\Longleftrightarrow\quad
  P\cup Z\text{ is feasible}.
\]
Constructing these vectors and maps is the main task in
Section~\ref{sec:certificates}. Once they are available, a simple linear-basis selection
provides the compression.

Process the partial selections in nondecreasing cost order and retain a selection only
when its vector is independent of those already retained. A discarded vector is a linear
combination of retained vectors of no greater cost. If a linear map is nonzero on the
discarded vector, it must be nonzero on at least one vector in this combination. Applied
to \(L_Z\), this gives a retained selection that works with the same completion \(Z\).
The retained family is chosen once and has size at most the dimension of the vector
space; the successful retained selection may depend on \(Z\). The following lemma states
this principle for arbitrary objects and linear maps.

\begin{lemma}\label{lem:basis-representatives}

Let \(\mathcal{A}\) be a finite family of objects, each with a cost \(c(A)\) and an associated
vector \(v(A)\) in a finite-dimensional vector space \(U\) over \(\mathbb{F}\). Process the
objects in nondecreasing order of costs, retaining an object exactly when its vector does not
belong to the span of the previously retained vectors. The resulting subfamily
\(\widehat{\mathcal{A}}\) has size at most \(\dim U\). Moreover, for every \(\mathbb{F}\)-linear
map \(L:U\to W\) and every \(A\in\mathcal{A}\) with \(L(v(A))\ne0\), there is an object
\(A'\in\widehat{\mathcal{A}}\) such that \(c(A')\le c(A)\) and \(L(v(A'))\ne0\).

\end{lemma}

\begin{proof}

The retained vectors are linearly independent, so \(|\widehat{\mathcal{A}}|\le\dim U\). Fix an
\(\mathbb{F}\)-linear map \(L:U\to W\) and an object \(A\in\mathcal{A}\) such that
\(L(v(A))\ne0\). If \(A\) is retained, take \(A'=A\). Otherwise, when \(A\) is processed, the
vector \(v(A)\) lies in the span of the previously retained vectors. Thus
\begin{equation}
  \label{eq:vector-expansion}
  v(A)=\sum^{t}_{j=1}\alpha_{j}v(A_{j}),\qquad \alpha_{j}\in\mathbb{F},
\end{equation}
where \(A_{1},\dots,A_{t}\) are previously retained objects. Since the objects are processed in
nondecreasing order of cost, \(c(A_{j})\le c(A)\) for every \(j\in[t]\). Applying \(L\) to
\eqref{eq:vector-expansion} gives
\[
  L(v(A))=\sum^{t}_{j=1}\alpha_{j}L(v(A_{j})).
\]
Since the left-hand side is nonzero, \(L(v(A_{j}))\ne0\) for some \(j\in[t]\). Taking
\(A'=A_{j}\) proves the claim.
\qed
\end{proof}

The target space \(W\) need not be finite-dimensional. In particular, it may be a
polynomial ring over \(\mathbb F\), viewed as an \(\mathbb F\)-vector space. In the
fault-tolerant application, \(v(P)\) will be the coefficient vector of a polynomial
\(G_P\), and \(L_Z\) will apply the symbolic substitution associated with \(Z\).
Although this substitution replaces variables by polynomials, it is linear over
\(\mathbb F\): it preserves sums and multiplication by field scalars. Thus the lemma
applies to the finite-dimensional coefficient space.

Here a nonzero output means a polynomial that is \emph{not identically zero}; we do not
assign numerical values to its remaining indeterminates. This distinction matters over
small finite fields, where a nonzero formal polynomial can vanish at every assignment
of field elements. The representative selection uses exact linear algebra on coefficient
vectors and requires no random evaluation.

Before constructing these polynomials, we illustrate the same principle with
simultaneous spanning in several linear matroids. In that application, a tensor product
combines the spanning tests into one linear functional, so one retained partial selection
works with a fixed completion in every matroid. The resulting lemma will also be used
for explicitly listed failure scenarios.

\begin{lemma}\label{lem:simultaneous-spanning}

Let \(M_{1},\dots,M_{m}\) be linear matroids on a common \(n\)-element ground set \(E\),
represented over the same field \(\mathbb{F}\), with respective ranks \(r_{1},\dots,r_{m}\). Let
\(w:E\to \mathbb{Z}_{\ge0}\) be a cost function and put \(R=\sum^{m}_{i=1}r_{i}\). A
minimum-cost set \(B\subseteq E\) satisfying \(\rank_{M_{i}}(B)=r_{i}\) for every \(i\in[m]\)
can be found using \(2^{\mathcal{O}(R)}n^{\mathcal{O}(1)}\) field operations after reading and reducing the
representations. Including this preprocessing, the running time is
\(2^{\mathcal{O}(R)}\operatorname{poly}(L_{\mathrm{in}})\) in the field-operation model, where
\(L_{\mathrm{in}}\) is the input encoding length.

\end{lemma}

\begin{proof}

Discard all rank-zero matroids, since every subset of \(E\) spans them. If no constraints
remain, return the empty set. We may therefore assume that every \(r_{i}\) is positive and
\(m\le R\). For each \(i\), retain \(r_{i}\) independent rows of its representing matrix,
obtaining a matrix \(A_{i}\) representing the same matroid \(M_{i}\). Fix a common ground-set
order \(e_{1},\dots,e_{n}\) and write \(a_{i}(e)\in\mathbb{F}^{r_{i}}\) for the column of \(e\)
in \(A_{i}\).

For a partial set \(P\) selected from a processed prefix, let
\(\mathbf p(P)=(p_1,\dots,p_m)\), where \(p_i=\rank_{M_i}(P)\), be its \emph{rank profile}.
We first compress an arbitrary family of partial sets with one fixed profile. The construction
has three ingredients: a vector encoding each partial span, a linear test for completing that
span, and a tensor product combining the tests for all matroids.

\smallskip\noindent\textbf{Encoding a partial span by its minors.}
For each \(i\in[m]\), choose an ordered subset
\(B_i(P)=(b_{i,1},\dots,b_{i,p_i})\) of \(P\) whose columns form a basis of its span in
\(A_i\). Define
\[
  v_i(P)=a_i(b_{i,1})\wedge\cdots\wedge a_i(b_{i,p_i})
  \in U_i:=\bigwedge^{p_i}\mathbb F^{r_i}.
\]
Concretely, \(v_i(P)\) is a vector with one coordinate for each \(p_i\)-element subset of
the rows: that coordinate is the corresponding \(p_i\times p_i\) minor of the chosen
basis matrix. Thus \(\dim U_i=\binom{r_i}{p_i}\). When \(p_i=0\), we use
\(v_i(P)=1\) in \(\bigwedge^0\mathbb F^{r_i}=\mathbb F\). Changing the ordered basis of
the same span multiplies all these coordinates by one nonzero scalar, which does not affect
any of the nonvanishing tests below.

\smallskip\noindent\textbf{Testing a completion by a determinant.}
Fix a set \(Z\) of unprocessed elements such that \(P\cup Z\) spans every matroid. For each
\(i\), extend \(B_i(P)\) to a basis of \(M_i\) using an ordered subset
\(Q_i=(z_{i,1},\dots,z_{i,r_i-p_i})\) of \(Z\). The subsets \(Q_i\) may differ between
matroids, but they are all taken from the same completion \(Z\). Identifying the top exterior
power \(\bigwedge^{r_i}\mathbb F^{r_i}\) with \(\mathbb F\), define
\[
  \lambda_i:U_i\longrightarrow\mathbb F,
  \qquad u\longmapsto
  u\wedge a_i(z_{i,1})\wedge\cdots\wedge a_i(z_{i,r_i-p_i}).
\]
This is a linear functional. In coordinates, appending the fixed columns of \(Q_i\) gives
a determinant that is a linear combination of the minors stored in \(u\), by Laplace
expansion. In particular,
\[
  \lambda_i(v_i(P))
  =\det[\,a_i(b_{i,1})\mid\cdots\mid a_i(b_{i,p_i})
       \mid a_i(z_{i,1})\mid\cdots\mid a_i(z_{i,r_i-p_i})\,]\ne0.
\]
If \(p_i=r_i\), then \(Q_i\) is empty and \(\lambda_i\) is just the identification of
the top exterior power with \(\mathbb F\). For any other partial set \(P'\) with the same
profile, \(\lambda_i(v_i(P'))\ne0\) implies that \(B_i(P')\cup Q_i\) is a basis, so
\(P'\cup Z\) spans \(M_i\). These witness subsets and functionals are used only to prove
preservation; the compression procedure does not need to know \(Z\).

\smallskip\noindent\textbf{Combining all spanning requirements.}
Applying the representative lemma separately in each \(U_i\) could produce different
retained partial sets for different matroids. To obtain one partial set that passes all the
tests, associate with \(P\) the tensor vector
\[
  \Phi(P)=\bigotimes_{i=1}^m v_i(P)
  \in U_{\mathbf p}:=\bigotimes_{i=1}^m U_i,
  \qquad
  D_{\mathbf p}:=\dim U_{\mathbf p}=\prod_{i=1}^m\binom{r_i}{p_i}.
\]
Its coordinates are all products formed by taking one coordinate from each \(v_i(P)\).
The tensor product of the functionals is a linear functional
\(\Lambda=\lambda_1\otimes\cdots\otimes\lambda_m:U_{\mathbf p}\to\mathbb F\) satisfying
\[
  \Lambda(\Phi(P))=\prod_{i=1}^m\lambda_i(v_i(P))\ne0.
\]
A product of field elements is nonzero exactly when every factor is nonzero. Thus this
single linear test enforces all the spanning requirements simultaneously.

For each rank profile, process the candidate partial sets in nondecreasing cost order and
retain a set precisely when its tensor vector is independent of the previously retained
vectors. This retains at most \(D_{\mathbf p}\) sets. By
Lemma~\ref{lem:basis-representatives}, the nonzero value of \(\Lambda(\Phi(P))\) guarantees
a retained set \(P'\) with \(w(P')\le w(P)\) and \(\Lambda(\Phi(P'))\ne0\).
Every factor \(\lambda_i(v_i(P'))\) is therefore nonzero, so \(B_i(P')\) together with
\(Q_i\) forms a basis of \(M_i\) for every \(i\). Consequently, the same retained set
\(P'\) works with the same completion \(Z\) in all matroids. This proves the required
compression guarantee within a fixed rank profile.

\smallskip\noindent\textbf{The dynamic program.}
We now use this compression procedure in a dynamic program that processes \(e_{1},\dots,e_{n}\)
in order. After each processed prefix, we maintain a retained family of partial sets for every
rank profile. Initially, only the empty set is retained, in profile \((0,\dots,0)\). When
processing the next element \(e\), generate the two successors \(P\) and \(P\cup\{e\}\) from
every retained partial set \(P\). Compute the rank profile of each successor and group all
successors with the same profile together. Apply the compression procedure to each group to
obtain the retained families for the new prefix. After processing all elements, return a
cheapest retained set with profile \((r_{1},\dots,r_{m})\). Such a set exists because \(E\)
spans every \(M_{i}\) and compression preserves successful completions.

\smallskip\noindent\textbf{Preserving an optimum.}
For correctness, we prove by induction on the processed prefix that whenever \(P\subseteq
\{e_{1},\dots,e_{j}\}\) and \(Z\subseteq\{e_{j+1},\dots,e_{n}\}\) satisfy that \(P\cup Z\) spans
every matroid, some retained set \(P'\) has \(w(P')\le w(P)\) and \(P'\cup Z\) spans every
matroid. The profiles of \(P\) and \(P'\) need not agree. The claim holds for the empty prefix.
Assume that it holds after processing \(j-1\) elements and let \(e=e_{j}\). If \(e\notin P\),
apply the inductive hypothesis to \(P\) with completion \(Z\). It gives a retained set
\(\widehat{P}\) of no greater cost whose skip successor has successful completion \(Z\). If
\(e\in P\), apply the hypothesis to \(P\setminus\{e\}\) with completion \(Z\cup\{e\}\). It gives
a retained set \(\widehat{P}\) with \(w(\widehat{P})\le w(P\setminus\{e\})\) such that
\(\widehat{P}\cup Z\cup\{e\}\) spans every matroid. Its take successor \(\widehat{P}\cup\{e\}\)
has cost at most \(w(P)\) and successful completion \(Z\). In either case, compression in the
successor's actual profile preserves a suitable set by the preceding argument. At the final
prefix, applying the invariant to an optimum with \(Z=\emptyset\) proves optimality of the
returned set.

\smallskip\noindent{\textbf{Bounding the retained families and running time.}}
The total number of retained sets at any prefix is at most
\[
  \begin{aligned}
    \sum_{p_1=0}^{r_1}\dots\sum_{p_m=0}^{r_m}D_{\mathbf p}
    &=\sum_{p_1=0}^{r_1}\dots\sum_{p_m=0}^{r_m}
      \left(\prod_{i=1}^{m}\binom{r_i}{p_i}\right)\\
    &=\prod_{i=1}^{m}\left(\sum_{p_i=0}^{r_i}\binom{r_i}{p_i}\right)
      =\prod_{i=1}^{m}2^{r_i}=2^R,\qquad \text{as }\sum_{i=1}^m r_i=R.
  \end{aligned}
\]
Thus each prefix generates at most \(2^{R+1}\) candidates. A skip successor inherits its
parent's basis columns. For a take successor, append the new column in each matroid where it
increases the rank. This gives the successor's rank profile and basis columns in
\(\operatorname{poly}(R)\) field operations. Each exterior coordinate is a minor of order at
most \(R\) and each tensor coordinate is a product of at most \(R\) exterior coordinates. Since
\(D_{\mathbf p}\le2^{R}\), forming the vectors and compressing the candidate families takes
\(2^{\mathcal{O}(R)}\) field operations per prefix and \(2^{\mathcal{O}(R)}n^{\mathcal{O}(1)}\) overall. Cost additions and
comparisons use integers of polynomial binary length. The selected sets or predecessor
information suffice to reconstruct an optimum. Reading and reducing the representations takes
polynomial input time.
\qed
\end{proof}

\subsection{A polynomial certificate for common completions}\label{sec:certificates}

We now construct the vectors and linear maps promised in
Section~\ref{sec:linear-representatives}. For each partial selection \(P\), we will build a
polynomial \(G_P\). A fixed future completion \(Z\) will determine a symbolic substitution
\(\ev_Z\), common to all partial selections in the same state, with the property
\[
  \ev_Z(G_P)\not\equiv0
  \quad\Longleftrightarrow\quad
  P\cup Z\text{ survives every allowed failure}.
\]
The coefficient vector of \(G_P\) will then be the vector to which we apply
Lemma~\ref{lem:basis-representatives}.

There can be many failure patterns. Instead of introducing a separate tensor factor for
each test, we will express all tests in the same polynomial ring and multiply them.
Multiplication increases the degree, but introduces no new variables. This is the source
of the eventual dimension bound: the ring has only \(\mathcal{O}(r(r+k))\) variables, irrespective
of the number of failure patterns. We first describe a completion by formal weighted sums
of its columns, then filter out failed future columns, and finally test the surviving
rank by a determinant. The product of these tests will enforce all failures together.

\smallskip\noindent\textbf{Fixing a state and its future slots.}
Fix a target solution size \(0\le s\le b=(k+1)r\) and a target vulnerable count
\(0\le s_V\le s\). A partial set \(P\) in the processed prefix has state
\((a_S,a_V,p)\), where
\[
  a_S=|P\cap S|,\qquad a_V=|P\cap V|,\qquad p=\rank_M(P).
\]
Its rank deficiency and the required numbers of future elements are
\[
  \begin{aligned}
    d&=r-p, & q_S&=s-s_V-a_S, & q_V&=s_V-a_V,\\
    && q&=q_S+q_V=s-a_S-a_V.
  \end{aligned}
\]
We consider only states with \(q_S,q_V\ge0\). A completion \(Z\) consists of unprocessed
elements with exactly \(q_S\) safe and \(q_V\) vulnerable elements.

To describe every such completion in the same variables, fix \(q\) future slots. The
first \(q_S\) are safe and the remaining \(q_V\) are vulnerable; write
\([q]_V=\{q_S+1,\dots,q\}\), interpreted as the empty set when \(q_V=0\).
Assign the safe elements of \(Z\) to the safe slots in ground-set order, and assign its
vulnerable elements to the vulnerable slots in the same way. Let \(z_j\) occupy slot \(j\).
This assignment depends only on \(Z\) and the state counts, so it is identical for every
partial set in the state. In particular, a fixed failed-slot set describes the same failed
future elements when we compare different partial sets.

\smallskip\noindent\textbf{Representing a completion by shared moments.}
We use \(d+k\) vector variables to stand for weighted sums of future columns. This number
is sufficient because a failure of at most \(k\) selected elements increases the current
rank deficiency by at most \(k\). Each future slot also receives a formal tag; these tags
will allow us to remove precisely the slots that fail. The variables are shared by all
failure patterns and by all partial sets in the state.

Introduce vector indeterminates \(X_{h}=(X_{h,1},\dots,X_{h,r})^{\mathsf T}\) for \(0\le h<d+k\)
and scalar tag indeterminates \(T_{1},\dots,T_{q}\). Let
\begin{equation}
  \label{eq:moment-ring}
 \mathcal R_{p,q}
  =\mathbb F[X_{h,\ell}:0\le h<d+k,\ 1\le\ell\le r]
     [T_1,\dots,T_q].
\end{equation}
This is the polynomial ring over \(\mathbb{F}\) in the moment coordinates and slot tags. In
particular, \(X_{h}\) is a vector in \(\mathcal{R}_{p,q}^{r}\), rather than a scalar polynomial.
All coordinates and tags are independent formal variables.

The polynomial certificate is constructed before the future columns are known. A completion
is used only to interpret its variables through the following substitution. This is the
linear map used in the preservation proof, not an operation that requires the algorithm
to enumerate completions.

For a completion \(Z\), introduce additional indeterminates \(u_{1},\dots,u_{q}\) and write
\(a_{z_{j}}\in \mathbb F^r\) for the column of \(z_j\). Define the \(\mathbb F\)-algebra
homomorphism
\[
  \ev_Z:\mathcal R_{p,q}\to\mathbb F[T_1,\dots,T_q,u_1,\dots,u_q]
\]
by fixing each tag and substituting
\begin{equation}
  \label{eq:moment-substitution}
  \ev_{Z}(X_{h,\ell})=\sum^{q}_{j=1}u_{j}T^{h}_{j}a_{z_{j},\ell},\qquad 0\le h<d+k,1\le\ell\le r.
\end{equation}
The map preserves field constants, addition and multiplication. Applied coordinatewise, it sends
\(X_{h}\) to \(\sum^{q}_{j=1}u_{j}T^{h}_{j}a_{z_{j}}\). These weighted sums are the moments of
the completion. The variables \(u_{j}\) are used only in this interpretation and do not belong
to \(\mathcal{R}_{p,q}\). We write \(G|_{Z}=\ev_{Z}(G)\) for a polynomial or polynomial vector
\(G\). If \(q=0\), each moment coordinate is sent to zero. All nonvanishing statements below
concern formal polynomials after substitution.

\smallskip\noindent\textbf{Filtering out failed future columns.}
Fix the future slots that fail. We want to remove their contributions from every weighted
sum without introducing new variables. A polynomial vanishing at exactly their formal
tags provides the required filter: multiplying the contribution of slot \(j\) by its
value at \(T_j\) kills a failed slot and leaves a nonzero formal weight on every survivor.

For a set \(J\subseteq[q]_{V}\) of failed future slots, define the filter polynomial and
filtered moments by
\begin{equation}
  \label{eq:filter-polynomial}
  \Fltr_{J}(t)=\prod_{j\in J}(t-T_{j})=\sum^{|J|}_{\nu=0}c_{J,\nu}t^{\nu},
\end{equation}
\begin{equation}
  \label{eq:filtered-moments}
  Y_{J,h}=\sum^{|J|}_{\nu=0}c_{J,\nu}X_{h+\nu},\qquad 0\le h\le d+k-|J|-1.
\end{equation}
Here \(c_{J,\nu}\in\mathbb{F}[T_{1},\dots,T_{q}]\) and each \(Y_{J,h}\) belongs to
\(\mathcal{R}_{p,q}^{r}\). When \(J=\emptyset\), the filter is 1 and \(Y_{\emptyset,h}=X_{h}\).
Substituting \eqref{eq:moment-substitution} into \eqref{eq:filtered-moments} gives
\begin{equation}
  \label{eq:filtered-substitution}
  Y_{J,h}|_{Z}=\sum^{q}_{j=1}u_{j}T^{h}_{j}\Bigl(\sum^{|J|}_{\nu=0}c_{J,\nu}T^{\nu}_{j}\Bigr)a_{z_{j}}=\sum_{j\in[q]\setminus J}u_{j}T^{h}_{j}\Fltr_{J}(T_{j})a_{z_{j}}.
\end{equation}
The failed terms vanish because \(\Fltr_{J}(T_{j})=0\) for \(j\in J\). For \(j\notin J\), the
polynomial \(\Fltr_{J}(T_{j})=\prod_{v\in J}(T_{j}-T_{v})\) is nonzero. Thus the filter removes
failed future columns while preserving a formal contribution from every survivor.

For example, take \(r=2\) and \(k=1\) and let the partial solution consist of one nonzero column
\(a\), so \(d=1\) and the available moments are \(X_{0},X_{1}\). Suppose that both future slots
are vulnerable and the first fails. Then \(\Fltr_{\{1\}}(t)=t-T_{1}\) and
\(Y_{\{1\},0}=X_{1}-T_{1}X_{0}\). Consequently, \(Y_{\{1\},0}|_{Z}=u_{2}(T_{2}-T_{1})a_{z_{2}}\)
and
\begin{equation}
  \label{eq:filter-example}
  \det[a\mid Y_{\{1\},0}]|_{Z}=u_{2}(T_{2}-T_{1})\det[a\mid a_{z_{2}}].
\end{equation}
This polynomial is nonzero exactly when the surviving future column completes the past span. The
factor \(T_{2}-T_{1}\) remains nonzero over \(\mathbb{F}_{2}\) because the tags are distinct
indeterminates.

We now define the rank test for a single failure pattern. Fix a partial set \(P\) in the state
and let \(I\subseteq P\cap V\) and \(J\subseteq[q]_{V}\) satisfy \(|I|+|J|\le k\). Choose an
ordered basis of the span of \(P\setminus I\) from its input columns and let \(A_{I}\) be the
resulting matrix. Put \(\delta_{I}=r-\rank_{M}(P\setminus I)\), so \(A_{I}\) has
\(r-\delta_{I}\) columns. Since deleting \(|I|\) elements decreases rank by at most \(|I|\), we
have \(\delta_{I}\le d+|I|\). Define
\begin{equation}
  \label{eq:failure-determinant}
  \Delta_{P,I,J}=\det[A_{I}|Y_{J,0}|\cdots |Y_{J,\delta_{I}-1}]\; \text{if}\; \delta_{I}>0,
  \qquad \Delta_{P,I,J}=1\; \text{if}\; \delta_{I}=0.
\end{equation}
The determinant has exactly \(r\) columns. All moments it needs are available, because its
largest moment index is \(\delta_{I}-1+|J|\le d+|I|+|J|-1\le d+k-1\). For \(\delta_{I}=0\), the
surviving past set already spans the matroid. Changing the ordered basis matrix \(A_{I}\)
multiplies a determinant with \(\delta_{I}>0\) by a nonzero field scalar, preserving its
vanishing under every completion substitution.

\begin{lemma}\label{lem:single-failure}

Fix a partial set \(P\) and a completion \(Z=(z_{1},\dots,z_{q})\) occupying the prescribed
slots. For every failure set \(F\subseteq(P\cup Z)\cap V\) with \(|F|\le k\), let \(I=F\cap P\)
and \(J=\{j\in[q]_{V}:z_{j}\in F\}\), so \(F=I\mathbin{\dot\cup}\{z_{j}:j\in J\}\). Then
\begin{equation}
  \label{eq:determinant-test}
  \Delta_{P,I,J}|_{Z}\not\equiv0\Longleftrightarrow \rank_{M}\bigl((P\setminus I)\cup(Z\setminus J)\bigr)=r,
\end{equation}
where \(Z\setminus J\) denotes deletion of the elements occupying slots in \(J\).

\end{lemma}

\begin{proof}

If \(\delta_{I}=0\), both sides hold. Assume that \(\delta=\delta_{I}>0\) and write
\(K=[q]\setminus J\). By \eqref{eq:filtered-substitution}, each substituted filtered column is a
linear combination of columns indexed by \(K\). If the surviving past and future columns have
rank less than \(r\), every \(r\times r\) determinant formed from their span vanishes, so
\(\Delta_{P,I,J}|_{Z}\equiv0\).
In particular, if fewer than \(\delta\) future columns survive, they cannot
fill the rank deficiency, and the determinant vanishes. This includes an empty completion
when the surviving past set does not span.

For the converse, group the multilinear determinant expansion by the \(\delta\) distinct future
columns it uses. Equivalently, apply the Cauchy--Binet identity to the filtered-column matrix.
For \(L=\{\ell_{1}<\cdots <\ell_{\delta}\}\subseteq K\), the expansion is
\begin{equation}
  \label{eq:determinant-expansion}
  \begin{aligned}
    \Delta_{P,I,J}|_Z
    &=\sum_{\substack{L\subseteq K\\ |L|=\delta}}
      \det[A_I\mid a_{z_{\ell_1}}\mid\cdots\mid a_{z_{\ell_\delta}}]\,
      \left(\prod_{j\in L}u_j\Fltr_J(T_j)\right)\\
    &\qquad\cdot\prod_{1\le a<b\le\delta}(T_{\ell_b}-T_{\ell_a}).
  \end{aligned}
\end{equation}
The last product is the Vandermonde determinant \(\det[T^{h}_{\ell_{a}}]_{1\le a\le\delta,0\le
h<\delta}\). If the survivors have rank \(r\), some subset \(L\) extends the basis columns of
\(A_{I}\) to a full basis, making its input-column determinant nonzero. Every filter factor and
Vandermonde factor in that term is also a nonzero polynomial. Moreover, distinct subsets \(L\)
have distinct monomials \(\prod_{j\in L}u_{j}\), so these terms cannot cancel. The expansion is
therefore nonzero, proving the equivalence over every field.
\qed
\end{proof}

To require robustness against all allowed failures with one common completion, multiply the
tests in the same polynomial ring:
\begin{equation}
  \label{eq:obstruction-product}
  G_P=\prod_{\substack{I\subseteq P\cap V,\ J\subseteq[q]_V\\ |I|+|J|\le k}}\Delta_{P,I,J}\in\mathcal R_{p,q}.
\end{equation}
Sharing the moment and tag variables ensures that every factor is interpreted using the same
completion substitution.

\begin{lemma}\label{lem:common-completion}

For every future completion \(Z\) with the prescribed safe and vulnerable counts,
\(G_{P}|_{Z}\not\equiv0\) if and only if \(P\cup Z\) survives every allowed failure of size at
most \(k\).

\end{lemma}

\begin{proof}

The sets \(P\) and \(Z\) lie in disjoint processed and unprocessed portions of the ground set.
Each allowed failure therefore decomposes uniquely into failed past elements \(I\subseteq P\cap
V\) and failed future slots \(J\subseteq[q]_V\) with \(|I|+|J|\le k\). Conversely, every pair in
\eqref{eq:obstruction-product} defines such a failure. Since \(\ev_{Z}\) is an algebra
homomorphism, \(G_{P}|_{Z}=\prod_{I,J}(\Delta_{P,I,J}|_{Z})\) over these pairs. The target
polynomial ring is an integral domain, so this product is nonzero exactly when every factor is
nonzero. Lemma~\ref{lem:single-failure} identifies this condition with full rank after every
allowed failure.
\qed
\end{proof}

The basis witnessing full rank may depend on the failure pattern, but the completion set \(Z\)
is common to all patterns. This is the condition needed to compress partial solutions: choosing
separate representatives for separate failures would not guarantee a single retained partial
solution that works with \(Z\) throughout.

\begin{lemma}\label{lem:coefficient-representatives}

Fix target counts \((s,s_{V})\) and a state \((a_{S},a_{V},p)\). Let \(\mathcal{P}\) be a finite
family of partial sets in that state. Express their polynomials \(G_{P}\) in a common monomial
basis and retain a cost-ordered linear basis of the coefficient vectors. The resulting subfamily
\(\widehat{\mathcal{P}}\) has size at most the dimension of the coefficient space. For every
\(P\in\mathcal{P}\) and every future completion \(Z\) with the prescribed counts, if \(P\cup Z\)
is feasible, then some \(P'\in\widehat{\mathcal{P}}\) satisfies \(w(P')\le w(P)\) and \(P'\cup
Z\) is feasible.

\end{lemma}

\begin{proof}

All partial sets in the state use the same ring \(\mathcal{R}_{p,q}\). Choose a finite monomial
basis containing every monomial in their polynomials and identify each polynomial with its
coefficient vector. Fix a completion \(Z\) for which \(P\cup Z\) is feasible. Its slot
assignment is common to the state, so \(\ev_{Z}\) defines one \(\mathbb{F}\)-linear map from
this coefficient space to the target polynomial ring, viewed as an \(\mathbb{F}\)-vector space.
By Lemma~\ref{lem:common-completion}, this map is nonzero on the vector of \(G_{P}\).
Lemma~\ref{lem:basis-representatives} gives a retained set \(P'\) with \(w(P')\le w(P)\) and
\(G_{P'}|_{Z}\not\equiv0\). Lemma~\ref{lem:common-completion} then gives feasibility of \(P'\cup
Z\). The size bound follows from linear independence of the retained vectors.
\qed
\end{proof}

We can now process the ground set while applying this compression after each transition. The
degree bound on the obstruction polynomials controls the number of retained partial sets.

\subsection{Dynamic program and running time}\label{sec:linear-dp}

The coefficient representatives let us compress after each take-or-skip transition.
We first describe the algorithm and show that it preserves a cheapest partial selection
for every successful completion. We then bound the number of shared variables, the degree
of the certificates, and the resulting coefficient dimension. These bounds control both
the number of retained selections and the work needed to compute them.

\begin{theorem}\label{thm:linear}

\textup{\textsc{Weighted Nonuniform Fault-Tolerant Basis}} on an explicitly represented
rank-\(r\) linear matroid is solvable deterministically in \(2^{\mathcal{O}(kr(r+k)\log(r+1))}n^{\mathcal{O}(1)}\)
field operations, apart from polynomial-time preprocessing of the input representation. Over
standard finite-field encodings and over the rationals, the algorithm has
\(f(r,k)\operatorname{poly}(L_{\mathrm{in}})\) bit complexity for a computable function \(f\).

\end{theorem}

\begin{proof}

For \(r=0\), return the empty set. For \(k=0\), compute a minimum-cost basis by the weighted
matroid greedy algorithm. Assume henceforth that \(r,k\ge1\). By Lemma~\ref{lem:sparsification}, every feasible instance has an optimum of size at most
\(b=(k+1)r\).

\smallskip\noindent\textbf{States and initialization.}
Enumerate all target pairs \((s,s_{V})\) with \(0\le s_{V}\le s\le b\). For a fixed pair,
process a ground-set order \(e_{1},\dots,e_{n}\), maintaining a retained family in every state
\((a_{S},a_{V},p)\) with \(0\le a_S\le s-s_V\), \(0\le a_V\le s_V\) and \(0\le p\le r\).
A retained set in state \((a_S,a_V,p)\) contains exactly \(a_S\) safe and \(a_V\) vulnerable
elements and has rank \(p\). The remaining safe and vulnerable counts determine its future
slots, as in Section~\ref{sec:certificates}.
Initially, retain only \(\emptyset\) in state \((0,0,0)\).

\smallskip\noindent\textbf{Transitions and compression.}
When processing an element \(e\), generate the skip successor \(P\) from each retained set
\(P\). Also generate \(P\cup\{e\}\) if its type counts respect the target: a safe element may be
taken if \(a_{S}<s-s_{V}\) and a vulnerable element if \(a_{V}<s_{V}\). Compute each successor's
actual rank, place it in the corresponding state and compress every candidate family by
Lemma~\ref{lem:coefficient-representatives}. Within a compression, all coefficient vectors use
the same variables and monomial basis.

\smallskip\noindent\textbf{Terminal test and output.}
At the final prefix, inspect the states with \((a_{S},a_{V})=(s-s_{V},s_{V})\) and accept a
retained set exactly when it survives every vulnerable failure of size at most \(k\). This is
checked by enumerating the failures and computing survivor ranks. Equivalently, these states
have \(q=0\), so the only completion is \(\emptyset\) and its substitution sends all moment
coordinates to zero. By Lemma~\ref{lem:common-completion}, acceptance is therefore equivalent to
\(G_{P}(0)\ne0\); formal nonvanishing of \(G_{P}\) alone is insufficient. Return a cheapest
accepted set over all target pairs, or report infeasibility if none is accepted. The selected
sets or suitable predecessor information allow reconstruction of the solution.

\smallskip\noindent\textbf{Correctness invariant.}
Fix a target pair \((s,s_V)\). For every processed prefix
\(E_j=\{e_1,\dots,e_j\}\), we maintain the following property. If
\(P\subseteq E_j\) and \(Z\subseteq E\setminus E_j\) meet the target counts,
\begin{equation}
  \label{eq:target-counts}
  |P\cap S|+|Z\cap S|=s-s_V,\qquad
  |P\cap V|+|Z\cap V|=s_V,
\end{equation}
and \(P\cup Z\) is feasible, then there is a retained set \(P'\subseteq E_j\) such that
\[
  \begin{gathered}
    |P'\cap S|=|P\cap S|,\qquad |P'\cap V|=|P\cap V|,\\
    w(P')\le w(P),\qquad P'\cup Z\text{ is feasible}.
  \end{gathered}
\]
The ranks of \(P\) and \(P'\) need not agree. Rank is used to group candidates for each
individual compression. The invariant across transitions preserves counts, cost, and the
successful completion. This distinction lets us place each successor in its actual rank
state, without assuming that adding the next element changes the ranks of two partial
sets in the same way.

The invariant holds at the empty prefix. Suppose that it holds before processing
\(e=e_j\), and fix \(P,Z\) for the new prefix.

\emph{Skip case.} If \(e\notin P\), apply the previous invariant to \(P\) and \(Z\).
It yields a retained set \(\widehat P\) of the same safe and vulnerable counts, of cost
at most \(w(P)\), for which \(\widehat P\cup Z\) is feasible. Its skip successor is a
candidate with the required counts and successful completion \(Z\).

\emph{Take case.} If \(e\in P\), apply the previous invariant to
\(P\setminus\{e\}\) and completion \(Z\cup\{e\}\). This gives a retained set
\(\widehat P\) with the same counts as \(P\setminus\{e\}\), with
\(w(\widehat P)\le w(P\setminus\{e\})\), and with
\(\widehat P\cup Z\cup\{e\}\) feasible. Taking \(e\) is permitted because the resulting
counts equal those of \(P\). The take successor \(\widehat P\cup\{e\}\) has cost at most
\(w(P)\) and successful completion \(Z\).

In either case, the candidate is assigned to its actual rank state.
Lemma~\ref{lem:coefficient-representatives} then preserves a candidate in that state of
no greater cost that works with the same \(Z\). Its safe and vulnerable counts are
unchanged. This proves the invariant.

At the final prefix, apply the invariant to an optimum with the fixed target counts and
\(Z=\emptyset\). It gives a retained feasible set of no greater cost, which the terminal
test accepts. Only feasible sets are accepted. By Lemma~\ref{lem:sparsification}, some
enumerated target pair contains a global optimum, so choosing the cheapest accepted set
over all pairs proves optimality. If no set is accepted, no feasible solution exists.

\smallskip\noindent\textbf{Bounding the coefficient dimension.}
We track three quantities: the number \(N\) of variables, an upper bound \(D\) on the total
degree, and the dimension \(H\) of the polynomials of degree at most \(D\). The last quantity
bounds the size of every retained family within a state.

\emph{Number of variables.} For a state with \(d=r-p\) and \(q=s-a_S-a_V\), the shared ring
contains \(d+k\) moment vectors, each with \(r\) coordinates, and \(q\) slot tags. Thus
\[
  N=r(d+k)+q\le r(r+k)+(k+1)r=\mathcal{O}(r(r+k)).
\]
The auxiliary variables \(u_j\) are used only after substituting a completion, so they do
not contribute to the certificate's coefficient dimension.

\emph{Degree of the product.} We first bound the number of determinant factors and then
the degree of each factor.
Each failure
pair selects at most \(k\) positions among the \(a_{V}+q_{V}=s_{V}\le b\) vulnerable past
elements and future slots. Hence the number of factors in \(G_{P}\) is at most
\begin{equation}
  \label{eq:failure-count}
  \sum_{j=0}^{k}\binom{s_V}{j}\le N_{\mathrm{fail}}:=\sum_{j=0}^{k}\binom{b}{j}\le(2er)^k.
\end{equation}
For the last inequality, put \(t=k/b\le1\). Since \(t^{j}\ge t^{k}\) for \(0\le j\le k\), the
binomial theorem gives
\begin{equation}
  \label{eq:binomial-bound}
  \sum_{j=0}^{k}\binom{b}{j}\le t^{-k}(1+t)^b\le(eb/k)^k\le(2er)^k,
\end{equation}
using \((1+t)^{b}\le e^{bt}=e^{k}\) and \(b/k\le2r\).

The coefficient \(c_{J,\nu}\) has tag degree \(|J|-\nu\), so an entry of \(Y_{J,h}\) has total
degree at most \(|J|+1\). A determinant factor has \(\delta_{I}\) filtered columns and constant
past-basis columns. Its degree is at most \(\delta_{I}(|J|+1)\le r(k+1)\). Consequently,
\(G_{P}\) has total degree at most \(D=r(k+1)N_{\mathrm{fail}}\).

\emph{Dimension of the coefficient space.} Use one fixed ordering of all monomials in the
state's \(N\) variables of total degree at most \(D\). Every certificate in the state has
its coefficient vector in this space, whose dimension is
\begin{equation}
  \label{eq:coefficient-dimension}
  H=\binom{N+D}{N}\le(D+1)^N,\qquad \log H=\mathcal{O}(kr(r+k)\log(r+1)).
\end{equation}
The inequality holds because each of the \(N\) exponents lies between \(0\) and \(D\).
Moreover, \(D\le r(k+1)(2er)^k\), and hence
\[
  \log(D+1)=\mathcal{O}\bigl(\log r+\log(k+1)+k\log(r+1)\bigr)
  =\mathcal{O}\bigl(k\log(r+1)\bigr),
\]
using \(r,k\ge1\). Multiplying by the bound on \(N\) proves the stated bound on
\(\log H\). For the overall running time, use the maximum of these dimension bounds
over all states.

\smallskip\noindent\textbf{Constructing and compressing the certificates.}
For each candidate \(P\), enumerate its at most \(N_{\mathrm{fail}}\) failure pairs. Obtain each
required \(A_{I}\) by exact linear algebra on the at most \(b\) input columns of \(P\setminus
I\). Compute the filter coefficients by multiplying the linear factors, form the filtered
moments and construct the determinant factors. Store polynomials as dense coefficient vectors in
a fixed ordering of the \(H\) monomials, including zeros for absent monomials. Every partial
product has total degree at most \(D\). Naive polynomial multiplication uses \(H^{\mathcal{O}(1)}\) field
operations, and each determinant has at most \(r!\) permutation terms. Constructing \(G_{P}\)
therefore uses \(N_{\mathrm{fail}}r!\operatorname{poly}(b,r,k)H^{\mathcal{O}(1)}\) field operations, with
all coefficients computed exactly.

\smallskip\noindent\textbf{Total field-operation count.}
There are \(\mathcal{O}(b^{2})\) target pairs and \(\mathcal{O}(b^{2}r)\) states per pair. Each state retains at
most \(H\) sets, each generating at most two successors. Thus the number of candidates per
prefix is polynomial in \(b,r,H\). Cost sorting and Gaussian elimination on their coefficient
vectors have polynomial overhead in these quantities. Terminal verification uses at most
\(N_{\mathrm{fail}}\) rank tests per candidate. All factors, including \(N_{\mathrm{fail}}\) and
\(r!\), are absorbed by \(2^{\mathcal{O}(kr(r+k)\log(r+1))}\). Processing the \(n\) prefixes gives the
claimed field-operation bound.

Exact cost additions, comparisons and sorting have polynomial overhead in the cost encoding and
the number of candidates.

\smallskip\noindent\textbf{Bit complexity.} It remains to justify the bit-complexity claim. Over \(\mathbb{Q}\), retain independent original
rows and choose every \(A_{I}\) from original columns. Suppose the resulting input entries have
at most \(L\) bits. A partial set contains at most \(b\) columns and hence at most \(rb\)
entries. The product of their denominators gives a common denominator with \(\mathcal{O}(rbL)\) bits. Each
determinant factor has degree at most \(r\) in the input entries. Since \(G_{P}\) has at most
\(N_{\mathrm{fail}}\) factors, every coefficient has a denominator dividing the
\(rN_{\mathrm{fail}}\)th power of this common denominator, whose bit length is
\(\mathcal{O}(r^{2}bN_{\mathrm{fail}}L)\). The number of terms contributing to a coefficient, their
degrees in the input entries and the integer coefficients introduced by the filters are bounded
by functions of \(r,k\). Both numerator and denominator lengths are therefore at most
\(f_{1}(r,k)(L+1)\) for a computable function \(f_{1}\).

Clear denominators in each coefficient vector before cost-ordered basis selection. Scaling a
vector by a nonzero scalar preserves every dependence test. The elimination matrices have at
most \(H\) columns, so standard determinant bounds show that fraction-free Gaussian elimination
has intermediate bit lengths bounded by a parameter-dependent multiple of \(L+1\). Input rank
computations and cost comparisons also admit exact polynomial-time implementations. Together
with the operation bound, this gives \(f(r,k)\operatorname{poly}(L_{\mathrm{in}})\) bit complexity
over \(\mathbb{Q}\). For a finite field in a standard exact encoding, each field operation has
polynomial bit complexity in the field encoding length, giving the same conclusion. No
bit-complexity claim is made for an unspecified black-box field representation.
\qed
\end{proof}

\subsection{Fixed finite fields}\label{sec:finite-fields}

Over a fixed finite field \(\mathbb F_Q\), the number of possible column directions is bounded in terms
of the rank. This gives a simpler way to obtain a small ground set: within each direction,
we retain only a few inexpensive elements. We can then search this reduced set using the
small-solution bound from Section~\ref{sec:oracle}.

Two nonloop elements are \emph{parallel} if their columns are nonzero scalar multiples
of one another. A parallel class therefore represents one one-dimensional subspace of
\(\mathbb F_Q^r\), which we call its \emph{direction}. For spanning, the contribution of
a class depends only on whether at least one of its selected elements survives. One selected
safe element guarantees this; if all selected elements in the class are vulnerable, then
\(k+1\) selected elements suffice to guarantee a survivor under every allowed failure.
Selecting more cannot provide further protection.

Accordingly, we retain a cheapest safe element and up to \(k+1\) cheapest vulnerable
elements per class. These are available choices, not elements that every solution must
select. In particular, a solution may select fewer vulnerable elements from a class and
rely on other directions if all of them fail. We keep both safety types because their
costs may differ, and we do not know which choice an optimum makes.

\begin{theorem}\label{thm:finite}

Let \(M\) be a rank-\(r\) matroid represented over \(\mathbb{F}_{Q}\), where \(Q\) is the field
size. \textup{\textsc{Weighted Nonuniform Fault-Tolerant Basis}} can be reduced, without
changing the optimum, to at most \((k+2)(Q^{r}-1)/(Q-1)\) nonloop elements. Consequently, for
every fixed \(Q\), the problem is solvable in \(2^{\mathcal{O}(kr^{2}+kr\log(k+2))}n^{\mathcal{O}(1)}\) time.

\end{theorem}

\begin{proof}

\smallskip\noindent\textbf{Constructing the reduced ground set.}
Delete all loops, since they contribute nothing to spanning and removing them cannot increase
the cost. Retain \(r\) independent original rows of the representation. Divide each nonzero
column by its first nonzero coordinate and group identical normalized columns into parallel
classes. In every class, retain a cheapest safe element if one exists and the \(k+1\) cheapest
vulnerable elements, or all vulnerable elements if fewer are present. Let \(W\) be the union of
these retained elements. We show that every feasible set can be replaced by a feasible subset of
\(W\) of no greater cost.

\smallskip\noindent\textbf{A class with a selected safe element.}
First consider a class containing a selected safe element. Keep one such element and delete all
other selected elements in the class. Under every allowed failure, the retained safe element
survives and spans the direction of all deleted columns. Moreover, every allowed failure of the
resulting set is also allowed for the original set. Thus deletion preserves the survivor span
and feasibility, and nonnegative costs prevent any cost increase. Replace the remaining safe
element by a cheapest safe element in the class. The replacement is safe and has the same
direction, so it again preserves every survivor span without increasing cost.

\smallskip\noindent\textbf{A class with only selected vulnerable elements.}
Now consider a class with no selected safe element. If more than \(k+1\) vulnerable elements are
selected, keep any \(k+1\) and delete the others. Every allowed failure leaves at least one
retained element in the class, so its direction remains present in the survivor span. Deleting
the remaining columns therefore preserves feasibility and cannot increase cost. Let \(t\le k+1\)
be the number of selected elements left in the class. Replace them by the \(t\) cheapest
vulnerable elements of the class, all belonging to \(W\). Fix a bijection between the old and
new selected elements in this class and leave selected elements outside it fixed. The bijection
preserves vulnerability and direction. Pulling back any allowed failure of the new set gives an
allowed failure of the old set of the same cardinality, and the corresponding survivors span the
same directions. Hence the replacement preserves feasibility and does not increase cost.

Applying these operations to all classes yields a feasible subset of \(W\) of no greater
cost. The construction retains at least one element of every nonloop parallel class, so
\(W\) spans the original matroid and \(\rank(M|W)=r\). Thus restricting the ground set to
\(W\) preserves the target rank. Conversely, every feasible set in the restriction is
feasible in the original instance, since its columns, costs, safety types, and target rank
are unchanged. The optimum, including infeasibility, is therefore preserved.

\smallskip\noindent\textbf{Bounding the number of retained elements.}
There are \(Q^{r}-1\) nonzero vectors in \(\mathbb{F}^{r}_{Q}\), and each one-dimensional
subspace contains exactly \(Q-1\) of them. Hence there are \((Q^{r}-1)/(Q-1)\) possible
directions. Since at most one safe element and \(k+1\) vulnerable elements are retained per
direction,
\[
  |W|\le (k+2)\frac{Q^r-1}{Q-1}.
\]
This reduction is valid for every finite field. Fixing \(Q\) makes the number of directions
single-exponential in \(r\), which yields the running-time bound below.

\smallskip\noindent\textbf{Searching the reduced ground set.}
For \(r=0\), return the empty set; for \(k=0\), compute a minimum-cost basis. Assume henceforth
that \(r,k\ge1\). By the earlier sparsification lemma, a feasible reduced instance has an
optimum of size at most \(b=(k+1)r\). Enumerate all subsets \(B\subseteq W\) with \(|B|\le b\).
For each, enumerate all \(F\subseteq B\cap V\) with \(|F|\le k\) and test whether
\(\rank_{M}(B\setminus F)=r\). Return a cheapest feasible candidate, or report infeasibility if
none exists.

\smallskip\noindent\textbf{Running time.}
For fixed \(Q\), we have \(\log(|W|+1)=\mathcal{O}(r+\log(k+2))\). The number of candidates is at most
\begin{equation}
  \label{eq:finite-candidates}
  \sum_{j=0}^{b}\binom{|W|}{j}\le\sum_{j=0}^{b}|W|^j\le(|W|+1)^{b+1}.
\end{equation}
Each candidate has at most \(N_{\mathrm{fail}}\le(2er)^{k}\) failure sets, by
\eqref{eq:failure-count}, and each survivor rank requires a polynomial number of field
operations in \(b,r\).
Since \(b\le2kr\), the logarithm of the candidate bound is
\[
  \mathcal{O}\bigl((b+1)\log(|W|+1)\bigr)
  =\mathcal{O}\bigl(kr^2+kr\log(k+2)\bigr).
\]
The failure tests contribute at most \(\mathcal{O}(k\log(r+1))\) to the exponent, which is absorbed
by this bound. We therefore obtain
\(2^{\mathcal{O}(kr^2+kr\log(k+2))}n^{\mathcal{O}(1)}\) time for fixed \(Q\). The constants may depend on
\(Q\). Normalizing and grouping columns and sorting elements by cost take polynomially many
operations. Reading the encoded input and accounting for the bit lengths of costs add
polynomial overhead under the conventions of Section~\ref{sec:preliminaries}.
\qed
\end{proof}

In particular, every fixed failure budget gives a quadratic exponent in the rank;
the two-failure case is one such specialization.

\begin{corollary}\label{cor:two-faults}

For every fixed finite field and every fixed failure budget \(k\ge0\),
\textup{\textsc{Weighted Nonuniform Fault-Tolerant Basis}} is solvable in
\(2^{\mathcal{O}(r^2)}n^{\mathcal{O}(1)}\) operations.

\end{corollary}

\begin{proof}

For \(r,k\ge1\) with \(k\) fixed, Theorem~\ref{thm:finite} has exponent
\(\mathcal{O}(kr^2+kr\log(k+2))=\mathcal{O}(r^2)\). The hidden constant may depend on the field and on
\(k\). For rank zero, the empty set is optimal; for \(k=0\), a minimum-cost basis is
optimal and can be found in polynomially many operations.
\qed
\end{proof}

\section{Explicit scenarios: a parameterization boundary}\label{sec:scenarios}

The failure model now changes. Instead of requiring survival after every deletion of at
most \(k\) vulnerable elements, we are given an explicit list of scenarios
\(F_1,\dots,F_m\subseteq E\). A selected set must span after each listed scenario is
applied separately. The list specifies the required failures; we do not additionally
require survival after unions of scenarios. This is the bulk-robust spanning model.

Two different measures of the list lead to different parameterized complexity results.
The maximum scenario size measures how many elements can fail in any one listed event;
the list length \(m\) measures how many such events must be considered. We prove hardness
for rank plus maximum scenario size, even when every scenario has size two. In contrast,
rank plus list length yields an FPT algorithm for oracle matroids and a
\(2^{\mathcal{O}(rm)}n^{\mathcal{O}(1)}\) field-operation algorithm for represented matroids when \(m\ge1\).
The positive results will reuse the flat-avoidance representatives of Section~\ref{sec:oracle}
and the simultaneous-spanning lemma of Section~\ref{sec:linear-representatives}.

\begin{problembox}{Weighted Explicit-Scenario Fault-Tolerant Basis}
Given a rank-\(r\) matroid \(M\) on \(E\), nonnegative integer costs \(w\)
and scenarios \(F_1,\dots,F_m\subseteq E\), find a minimum-cost \(B\subseteq E\)
with \(\rank_M(B)=r\) and \(\rank_M(B\setminus F_i)=r\) for all \(i\in[m]\),
or report infeasibility.
\end{problembox}
\begin{problembox}{Explicit-Scenario Size-Budget Decision}
Given \(M\), the scenarios and a size budget \(\beta\), decide whether
a set satisfying the same rank conditions has size at most \(\beta\).
\end{problembox}

The no-failure constraint can equivalently be viewed as the implicit scenario
\(F_{0}=\emptyset\), which we do not count in \(m\). When \(m\ge1\), it already follows from any
listed scenario constraint by rank monotonicity. When \(m=0\), the problem is minimum-weight
spanning and is solved by a minimum-weight basis. Scenarios may contain elements outside the
selected set; such elements have no effect on its surviving rank.
The size-budget formulation is the decision version with unit costs: it asks whether the
optimum cardinality is at most \(\beta\). We use this restricted formulation for the
hardness result and allow arbitrary nonnegative costs in the algorithms.

\subsection{Hardness for two-element scenarios}\label{sec:scenario-hardness}

The reduction encodes an independent set using failures that are graph edges. In a uniform
matroid of rank \(r\), a selected set of size \(r+1\) tolerates a two-element scenario
exactly when the scenario contains at most one selected element. Requiring this for every
edge is precisely the independent-set condition.

We set \(r=\ell-1\), where \(\ell\) is the desired independent-set size, and impose the
budget \(r+1=\ell\). To force a feasible set within this budget to have exactly \(r+1\)
elements, we add a universal vertex. The resulting graph has no isolated vertices, so any
selection of only \(r\) elements loses rank under some edge scenario. An independent set
of size \(\ell\ge2\) cannot use the universal vertex, so the correspondence with the
original graph is preserved. The proof below treats the small exceptional instances
explicitly.

\begin{theorem}\label{thm:scenario-hardness}
The explicit-scenario size-budget problem is W[1]-hard parameterized by
\[
  r+\max_{i}|F_{i}|,
\]
even on uniform matroids, when every listed scenario has size exactly two and the solution
budget is \(r+1\).

\end{theorem}

\begin{proof}

We give a parameterized reduction from \textup{\textsc{Independent Set}}, parameterized by the
required independent-set size \(\ell\), which is W[1]-hard \cite{cygan}.

\smallskip\noindent\textbf{Exceptional instances.}
For \(\ell=0\), the source instance is a yes-instance. For \(\ell=1\),
it is a yes-instance exactly when \(V(G)\ne\emptyset\). These cases can therefore be
decided directly and mapped to the fixed instances below. For a yes-instance, output the uniform
matroid \(U_{1,\{a,b,c\}}\) with scenarios \(\{a,b\}\) and \(\{b,c\}\) and budget two. The set
\(\{a,c\}\) is feasible, since it retains one element after either scenario. For a no-instance,
output the uniform matroid \(U_{1,\{a,b\}}\) with the single scenario \(\{a,b\}\) and budget
two. This instance has no feasible solution, since the scenario deletes the entire ground set.
Both output instances have rank one, two-element scenarios and budget \(r+1=2\).

Instances with \(\ell>|V(G)|\) are no-instances and are mapped to the same fixed no-instance.
Henceforth, assume \(2\le\ell\le|V(G)|\).

\smallskip\noindent\textbf{The construction.}
Let \(G^{+}\) be obtained from \(G\) by adding a new vertex \(z\) adjacent to every vertex of
\(G\). Define
\[
  M=U_{\ell-1,V(G^{+})},\qquad r=\ell-1.
\]
Thus, a set spans \(M\) if and only if it contains at least \(r\) elements. For each edge
\(uv\in E(G^{+})\), include the scenario \(\{u,v\}\) and set the solution budget to
\(r+1=\ell\). We prove that \((G,\ell)\) is a yes-instance if and only if the resulting instance
admits a feasible solution.

\smallskip\noindent\textbf{An independent set gives a feasible selection.}
Suppose first that \(G\) contains an independent set \(I\) of size \(\ell\). Since \(|I|=r+1\),
the set \(I\) spans \(M\) and satisfies the budget. Moreover, \(I\) is independent in \(G^{+}\),
so every listed scenario \(F\) contains at most one element of \(I\). Consequently,
\[
  |I\setminus F|\ge|I|-1=r,
\]
and hence \(I\setminus F\) spans \(M\). Thus \(I\) is feasible.

\smallskip\noindent\textbf{A feasible selection gives an independent set.}
Conversely, let \(S\) be a feasible solution with \(|S|\le r+1\). We first show that
\(|S|=r+1\). Feasibility implies that \(S\) spans \(M\), so \(|S|\ge r\). If \(|S|=r\), choose
any \(x\in S\), which is possible because \(r\ge1\). The graph \(G^+\) has no isolated vertices: every original vertex is adjacent
to \(z\), and \(z\) has a neighbor because \(|V(G)|\ge\ell\ge2\). Hence some listed
scenario \(F\) contains \(x\). Then
\[
  |S\setminus F|\le r-1,
\]
contradicting the requirement that \(S\setminus F\) span \(M\). Therefore \(|S|=r+1=\ell\).

Next, no listed scenario can be contained in \(S\): if \(F\subseteq S\), then
\[
  |S\setminus F|=|S|-2=r-1,
\]
again contradicting feasibility. Since the listed scenarios are precisely the edges of
\(G^{+}\), the set \(S\) is independent in \(G^{+}\). As \(z\) is adjacent to every other vertex
and \(|S|=\ell\ge2\), we have \(z\notin S\). Hence \(S\subseteq V(G)\) is an independent set of
\(G\) of size \(\ell\).

\smallskip\noindent\textbf{Parameter bound.}
The construction takes polynomial time and every scenario has size exactly two. For
\(2\le\ell\le|V(G)|\), the output parameter is
\[
  r+\max_{i}|F_{i}|=(\ell-1)+2=\ell+1.
\]
In the exceptional cases, the output parameter is three. Thus the reduction is parameterized and
the claimed W[1]-hardness follows.
\qed
\end{proof}

\subsection{An algorithm for a bounded number of scenarios}\label{sec:few-scenarios}

The hardness reduction uses one scenario per edge, so the list can be long even when every
scenario has size two. We now bound the number \(m\) of scenarios together with the rank
\(r\), without bounding the size of an individual scenario.

The two representation models admit different algorithms. For oracle matroids, we group
elements according to which scenarios delete them and apply the flat-avoidance lemma
within each group. For represented matroids, we turn each scenario into a separate linear
matroid and apply the simultaneous-spanning algorithm. Both constructions preserve the
cost of an optimum.

\begin{theorem}\label{thm:few-scenarios}
On independence-oracle matroids,
the weighted explicit-scenario problem is fixed-parameter tractable
parameterized by \(r+m\). On represented linear matroids it is solvable in
\(2^{\mathcal{O}(rm)}n^{\mathcal{O}(1)}\) field operations after polynomial preprocessing
of the input representation. The case \(m=0\) is polynomial-time solvable.

\end{theorem}

\begin{proof}

\smallskip\noindent\textbf{Boundary cases and feasibility.}
The case \(r=0\) has the empty optimum, and the case \(m=0\) is minimum-weight spanning, so
assume \(r,m\ge1\). Delete loops and compute \(\rank_{M}(E\setminus F_{i})\) for each scenario.
If one of these ranks is below \(r\), even the whole ground set is infeasible, so reject.
If every test succeeds, \(E\) itself is feasible. Thus these tests decide feasibility;
the remaining task is to find a minimum-cost feasible selection. We give an oracle
algorithm and then a faster algorithm for an explicit linear representation.

\smallskip\noindent\textbf{A small optimum for the oracle algorithm.}
A feasible selection contains a surviving basis for each listed scenario. Taking the union
of one such basis per scenario gives a small feasible subset; nonnegative costs then give
the same size bound for some optimum.
To bound the size of an optimum, let \(B\) be a feasible set. For every \(i\in[m]\), choose a basis
\(I_{i}\subseteq B\setminus F_{i}\) of \(M\), and put \(C=\bigcup^{m}_{i=1}I_{i}\). Then
\(C\subseteq B\) with \(|C|\le mr\), and \(C\setminus F_{i}\) contains the full basis \(I_{i}\)
for every \(i\). Thus \(C\) is feasible. Nonnegative costs imply \(w(C)\le w(B)\), so there is
an optimum with at most
\[
  b=mr
\]
selected elements.

\smallskip\noindent\textbf{Grouping elements by failure signatures.}
In the safe/vulnerable model, an exchange preserves the safety type of the replaced
element. Here we preserve its membership in every listed scenario. This ensures that the
old and new elements either both fail or both survive under each scenario.
For oracle matroids, define the failure signature of \(e\in E\) by
\[
  \sigma(e)=\bigl(1[e\in F_1],\dots,1[e\in F_m]\bigr)\in\{0,1\}^m,
\]
that is, for each \(i\in[m]\),
\[
  \sigma_i(e)=
  \begin{cases}
    1 & \text{if }e\in F_i,\\
    0 & \text{if }e\notin F_i.
  \end{cases}
\]
A signature class consists of all elements with a fixed vector \(\sigma(e)\); there are
at most \(2^m\) such classes. For each nonempty class \(A\), apply
Lemma~\ref{lem:flat-avoidance} to the subset \(A\) of the ambient matroid \(M\), with
\(t=m\) and reserve \(b=mr\). The value \(t=m\) will cover all deficient survivor
hyperplanes arising from the listed scenarios. The reserve ensures that a replacement can
be chosen outside the other at most \(b-1\) selected elements. Let \(W\) be the union of
the resulting cores.
Since there are at most \(2^{m}\)
signature classes,
\begin{equation}
  \label{eq:signature-core-size}
  |W|\le2^{m}mr\sum^{r-1}_{j=0}m^{j}.
\end{equation}
In particular, this size depends only on \(r+m\).

\smallskip\noindent\textbf{Exchanging an element into the core.}
We show that \(W\) contains an optimum. Let \(B\) be an optimum with \(|B|\le b\), choose \(x\in
B\setminus W\), and put \(Y=B\setminus\{x\}\). For each scenario with \(x\notin F_{i}\) and
\(\rank(Y\setminus F_{i})<r\), feasibility of \(B\) implies
\[
  \rank(Y\setminus F_{i})=r-1,\qquad x\notin H_{i}:=\cl_{M}(Y\setminus F_{i}).
\]
Thus the distinct deficient hyperplanes form a family of size at most \(m\), and every
member avoids \(x\). Since \(|Y|\le b-1\), Lemma~\ref{lem:flat-avoidance} applies within
the signature class of \(x\), with forbidden set \(Y\). It returns
\(y\in W\setminus Y\) of the same signature, with \(w(y)\le w(x)\), outside every
hyperplane in this family. The lemma also covers the case when the family is empty.
Put \(B'=Y\cup\{y\}\).

\smallskip\noindent\textbf{Preserving every scenario.}
Fix a scenario \(F_{i}\). Since \(\sigma(y)=\sigma(x)\), either both \(x\) and \(y\) belong to
\(F_{i}\) or neither does. If \(x\in F_{i}\), then \(B'\setminus F_{i}=B\setminus F_{i}\), which
spans by feasibility of \(B\). Suppose now that \(x\notin F_{i}\). Since \(B\setminus
F_{i}=(Y\setminus F_{i})\cup\{x\}\) spans and adding one element increases rank by at most one,
\(Y\setminus F_{i}\) has rank either \(r\) or \(r-1\). In the former case, it already spans. In
the latter case, its closure is one of the deficient hyperplanes \(H_{i}\) constructed above.
The replacement element \(y\) was chosen outside all these hyperplanes, so adding \(y\)
increases the rank to \(r\). As \(y\notin F_{i}\), we conclude that \(B'\setminus
F_{i}=(Y\setminus F_{i})\cup\{y\}\) spans in both cases. Thus \(B'\) is feasible. Moreover,
\(w(y)\le w(x)\) gives \(w(B')\le w(B)\), and replacing \(x\notin W\) by \(y\in W\setminus Y\)
reduces the number of selected elements outside \(W\) by exactly one. After at most \(b\) such
exchanges there is an optimum entirely inside \(W\).

\smallskip\noindent\textbf{Searching the core.} Enumerate all subsets of \(W\) of size at
most \(b\), test their rank after each listed scenario, and choose a cheapest feasible
candidate. The size bound \eqref{eq:signature-core-size}, the construction bound in
Lemma~\ref{lem:flat-avoidance}, and greedy rank computation give \(f(r,m)n^{\mathcal{O}(1)}\)
independence queries and elementary operations for a computable function \(f\).

\smallskip\noindent\textbf{Linear representations: simultaneous spanning.}
With a linear representation, we can encode each listed scenario as a spanning constraint
on the same selected set. Zeroing a failed column makes it a loop while keeping its ground-set
label. This preserves a common ground set and a common cost function across all scenarios.
For a represented linear matroid, take each scenario \(F_i\) in turn and replace its
columns in the input representation by zero columns, obtaining a represented matroid \(M_{i}\) on the same
ground set. For every \(B\subseteq E\),
\[
  \rank_{M_{i}}(B)=\rank_{M}(B\setminus F_{i}).
\]
The initial feasibility test ensures that every \(M_{i}\) has rank \(r\). All \(M_{i}\) are
represented over the same field, so Lemma~\ref{lem:simultaneous-spanning} applies with \(R=mr\).
It finds a minimum-cost set spanning every \(M_{i}\) in \(2^{\mathcal{O}(rm)}n^{\mathcal{O}(1)}\) field operations.
By the rank equality above, this set is optimal for the scenario instance.
For \(m\ge1\), satisfying these spanning constraints also implies the no-failure
constraint by rank monotonicity, so no additional matroid for \(F_0=\emptyset\) is needed.
\qed
\end{proof}

\section{Transversal matroids and gammoids: one-failure hardness}\label{sec:gammoids}

We return to the failure-budget model. In this section all ground elements are vulnerable,
and a selected set must span after deletion of any one of its elements. Even this
one-failure case is hard on explicitly presented transversal matroids and on a restricted
class of gammoids.

The parameter here is different from those in the preceding sections. If \(\beta\) is the
solution-size budget and \(r\) is the matroid rank, the \emph{excess budget}
\(\beta-r\) counts how many elements beyond a basis may be selected. We will obtain
W[2]-hardness for this parameter. This is compatible with fixed-parameter tractability in
\(r+k\): the reduction keeps \(k=1\) and controls \(\beta-r\), but allows \(r\) to grow
with the source instance.

A transversal matroid is specified by a bipartite graph with ground elements on one side and
targets on the other. A subset of ground elements is independent if a matching covers all its
elements. Thus spanning a rank-\(r\) transversal matroid means that the selected elements admit
a matching of size \(r\).

The following NP-hardness construction is due to Adjiashvili, Bindewald, and Michaels in their
study of robust assignments with vulnerable nodes \cite[Section~4.1]{abm}. We include the proof to express it
in matroid language, record its direct gammoid interpretation, and make explicit the
parameterized consequence for the solution budget above rank. The NP-hardness and the underlying
\textup{\textsc{Set Cover}} gadget are not new claims here.

The construction has two kinds of ground elements. A pair of baseline elements for each
universe element is forced into every robust selection. Additional elements correspond
to input sets, and they repair a failed baseline element exactly when the selected sets
cover the corresponding universe element. This gives an exact relation between the two
optima, from which the excess-budget consequence follows.

\begin{theorem}\label{thm:one-failure}
The size-budget decision version of \textup{\textsc{Fault-Tolerant Basis}} is NP-complete on
explicitly presented transversal matroids already for \(k=1\) \cite{abm}. The same holds for
gammoids whose presenting digraph consists only of arcs from source vertices to ground-set
vertices. Moreover, the same construction gives W[2]-hardness parameterized by solution budget
minus rank, already with \(k=1\).

\end{theorem}

\begin{proof}

\smallskip\noindent\textbf{The Set Cover construction.}
We give the construction of \cite{abm} and its parameter accounting. Consider a
\textup{\textsc{Set Cover}} instance with universe \(U=\{u_1,\dots,u_m\}\), sets
\(A_1,\dots,A_s\subseteq U\) and budget \(q\). We may assume \(m\ge1\). Introduce two targets
\(u_{i},g_{i}\) for every \(i\in[m]\), two ground elements \(a_{i},d_{i}\) for every such \(i\),
and one ground element \(c_{j}\) for every input set \(A_{j}\). The neighborhoods of the ground
elements are
\[
  N(a_{i})=\{u_{i},g_{i}\},\qquad N(d_{i})=\{g_{i}\},\qquad N(c_{j})=\{u\in U:u\in A_{j}\}.
\]
All ground elements are vulnerable. Set \(k=1\) and the solution-size budget
\(\beta=2m+q\). The matroid has rank exactly \(2m\): there are
\(2m\) targets, and matching \(a_{i}\) to \(u_{i}\) and \(d_{i}\) to \(g_{i}\) for all \(i\)
saturates them all. We call this the baseline matching.

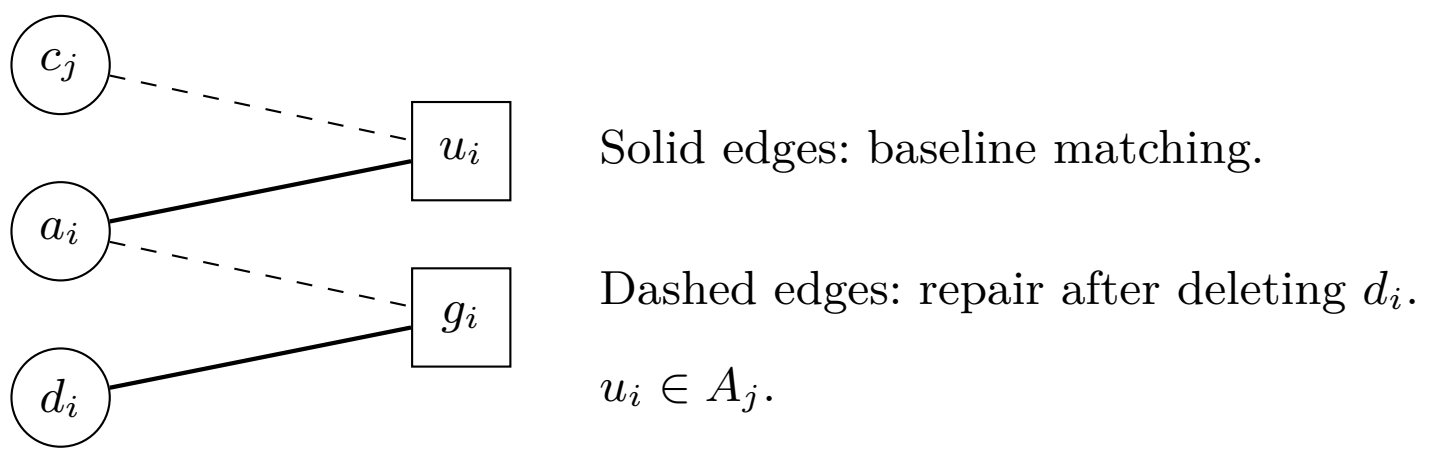
\begin{figure}[tb]
\centering
\begin{tikzpicture}[
  ground/.style={circle,draw,minimum size=6.5mm,inner sep=1pt},
  target/.style={rectangle,draw,minimum size=6.5mm,inner sep=1pt},
  x=1cm,y=1cm
]
\node[ground] (c) at (0,1.1) {\(c_j\)};
\node[ground] (a) at (0,0) {\(a_i\)};
\node[ground] (d) at (0,-1.1) {\(d_i\)};
\node[target] (u) at (2.65,.53) {\(u_i\)};
\node[target] (g) at (2.65,-.57) {\(g_i\)};
\draw[thick] (a) -- (u);
\draw[thick] (d) -- (g);
\draw[dashed] (c) -- (u);
\draw[dashed] (a) -- (g);
\node[anchor=west,font=\small] at (3.45,.53) {Solid edges: baseline matching.};
\node[anchor=west,font=\small] at (3.45,-.40) {Dashed edges: repair after deleting \(d_i\).};
\node[anchor=west,font=\small] at (3.45,-1.05) {\(u_i\in A_j\).};
\end{tikzpicture}
\caption{The local part of the robust-assignment construction.
After deleting $d_i$, match $a_i$ to $g_i$ and $c_j$ to $u_i$.
All baseline edges at other indices remain unchanged.
Only one failure is repaired at a time.}
\label{fig:repair}
\end{figure}

\smallskip\noindent\textbf{Every baseline element is mandatory.}
Let \(B\) be one-fault tolerant. Since \(B\) spans
before failures, it contains at least one of \(a_{i},d_{i}\): these are the only neighbors of
\(g_{i}\), and a size-\(2m\) matching must cover every target. If exactly one were selected,
deleting that element would leave \(g_{i}\) with no selected neighbor. Hence both \(a_{i}\) and
\(d_{i}\) belong to \(B\) for every \(i\).

\smallskip\noindent\textbf{The additional elements encode a cover.}
The selected set-elements must also form a cover. Let \(C=\{j:c_j\in B\}\). Delete \(a_i\). Any
surviving size-\(2m\) matching must match \(g_{i}\) to \(d_{i}\). It must also cover \(u_{i}\),
whose only remaining possible selected neighbors are set-elements \(c_{j}\) with \(u_{i}\in
A_{j}\). Therefore such an index \(j\in C\) exists for every \(i\), so the sets indexed by \(C\)
cover \(U\).

\smallskip\noindent\textbf{A cover repairs every single failure.}
Conversely, let \(C\) index a set cover and select all
baseline elements together with the corresponding \(c_j\). The baseline matching proves
spanning without failures.

There are three possible types of a failed selected element.
\smallskip\noindent\emph{A set-element \(c_j\) fails.}
Deleting a selected \(c_j\) leaves the baseline matching intact.

\smallskip\noindent\emph{A baseline element \(a_i\) fails.}
Retain every baseline pair at indices different from \(i\), match \(d_i\) to \(g_i\),
and match a selected covering \(c_j\) to \(u_i\).

\smallskip\noindent\emph{A baseline element \(d_i\) fails.}
Retain every baseline pair at indices different from \(i\), match \(a_i\) to \(g_i\),
and match a selected covering \(c_j\) to \(u_i\).

In the last two cases the covering element \(c_j\) survives, because the unique failure
is a baseline element. The modified edges form a matching: the baseline matching uses no
set-elements, and the unchanged edges cover only targets at other indices. Each repair
therefore gives a matching of size \(2m\). Different failures may use different matchings;
only one failure must be repaired at a time. Figure~\ref{fig:repair} illustrates the local
repair.

\smallskip\noindent\textbf{Optimum identity and excess-budget hardness.}
It follows that there is a robust set of size at most \(2m+q\) if and only if the source
instance has a set cover of size at most \(q\). Equivalently, on feasible instances,
\[
  \OPT_{\mathrm{FTB}}=2m+\OPT_{\mathrm{SetCover}}.
\]
The construction has polynomial size.
The \(2m\) forced baseline elements account for the rank, and every additional selected
element corresponds to one input set. Thus the output parameter is exactly
\[
  \beta-r=(2m+q)-2m=q.
\]
The W[2]-hardness of \textup{\textsc{Set Cover}} parameterized by its
budget \cite{cygan} therefore gives the asserted parameterized consequence.

\smallskip\noindent\textbf{The same construction as a gammoid.}
For the gammoid presentation, use the convention that a gammoid is represented by a digraph, a
set of sources, and a ground set of terminal vertices; independence means that vertex-disjoint
paths link distinct sources to the chosen ground elements. Direct every edge of the bipartite
presentation from its target to its ground element, and take all targets as sources. Every
source-to-ground path has one edge, so a linkage is exactly a matching in the transversal
presentation. This gives the claimed restricted gammoid instances. Under the equivalent
convention of paths from ground elements to sinks, reverse all arcs.

\smallskip\noindent\textbf{Membership in NP.}
It remains to show membership in NP. For \(k=1\), a proposed selected set can be checked by computing its
rank and its rank after deletion of each selected element. In a transversal presentation these
are polynomial-time matching computations; in a gammoid presentation they are polynomial-time
vertex-capacitated flow computations. There are at most \(|B|+1\) checks, each polynomial in the
explicit presentation size. This proves NP membership and completes the theorem.
\qed
\end{proof}

\section{Conclusion}\label{sec:conclusion}

We developed two forms of compression for minimum-cost fault-tolerant spanning with safe
and vulnerable elements. For independence-oracle matroids, a bounded set of original
elements contains an optimum. The proof combines weighted flat-avoidance representatives
with a bound on small cocircuits through a prescribed element. For represented linear
matroids, we compress families of partial selections: every successful future completion
remains successful for a retained selection of no greater cost. Shared-variable polynomial
certificates ensure that one retained selection works with the same completion under all
allowed failures. Over fixed finite fields, a simpler reduction retains a bounded number
of inexpensive elements in each parallel class.

For explicitly listed failures, the two measures of the scenario list lead to different
complexity results. Rank together with the number of scenarios yields fixed-parameter
algorithms, using either failure signatures or simultaneous spanning. Rank together with
the maximum scenario size gives W[1]-hardness, even for uniform matroids and two-element
scenarios. We also expressed the robust-assignment construction of
Adjiashvili, Bindewald, and Michaels~\cite{abm} in matroid language and derived its
W[2]-hardness consequence for solution budget minus rank on transversal matroids and
restricted gammoids, already for one failure.

The main algorithmic question left by our results is whether the parameter dependence of
common-completion compression can be improved over arbitrary fields. Our algorithm retains
a basis of coefficient vectors in the space of all monomials up to the certificate's
degree bound. Can the structure of these certificates yield a smaller space, or a way to
select representatives without expanding their full coefficient vectors, while preserving
the same deterministic guarantee and optimal cost?

\begin{credits}
\subsubsection{\ackname} 
ChatGPT was used to assist with language editing, organization, and clarity of exposition during manuscript preparation. AI was not used to develop scientific results or mathematical proofs, except for the W[1]-hardness result for explicit failure scenarios, which was developed with AI assistance. The authors independently examined and verified all mathematical statements, proofs, and arguments, including the AI-assisted hardness result. The authors take full responsibility for the correctness, originality, and scientific content of the manuscript.
\end{credits}

\end{document}